\documentclass[a4paper,11pt,final]{article}
\usepackage[titletoc,toc,title]{appendix}
\usepackage{fullpage} 
\usepackage{xcolor}

\makeatletter
	\renewcommand*{\@fnsymbol}[1]{\ensuremath{\ifcase#1\or\dagger\else\@ctrerr\fi}}
\makeatother

\usepackage{amsmath}
\usepackage{amssymb,wasysym}
\usepackage{dsfont}
\usepackage{upgreek}
\usepackage{mathtools}
\usepackage{enumitem}

\usepackage{siunitx}
\DeclareSIUnit\Molar{\textsc{m}}

\usepackage[version=4,arrows=pgf-filled]{mhchem}

\usepackage[outdir=./Figures/]{epstopdf}
\usepackage{graphicx}
\graphicspath{{./},{./Figures/}}
\usepackage[aboveskip=1pt,
            labelfont=bf,
            labelsep=period,
            justification=justified,
            singlelinecheck=off]{caption}
\usepackage[position=top,labelformat=simple]{subcaption}
\usepackage[export]{adjustbox}

\usepackage{natbib}
\allowdisplaybreaks

\usepackage{color}
\usepackage{xcolor}

\usepackage{listings}
\usepackage{hyperref}

\usepackage{tabularx}
\usepackage{longtable}
\let\oldhline\hline
\renewcommand{\hline}{\oldhline\noalign{\vskip 0.5ex}}
\usepackage{floatrow}
\usepackage[textsize=tiny,obeyDraft]{todonotes}

\newcommand{\der}[2]{\frac{\mathrm{d}#1}{\mathrm{d}#2}}
\newcommand{\rpar}[1]{\left(#1\right)}
\newcommand{\Hill}[3]{\ensuremath{\mathcal{H}_{#1}\left(#2,#3\right)}}

\newcommand{\dd}[1]{\ensuremath{\mathrm{d}#1}}
\renewcommand{\vec}[1]{\ensuremath{\mathbf{#1}}}
\def\ip3{\ce{IP3}}
\def\dag{\ce{DAG}}
\def\pip2{\ce{PIP2}}
\def\ca{\ce{Ca^2+}}
\def\plcd{PLC$\updelta$}
\def\plcb{PLC$\upbeta$}
\def\3k{\ip3\-3K}
\def\5p{IP-5P}
\def\cpkc{\ce{cPKC^*}}
\def\gchi{$G$-$ChI$}

\newcommand*{\tabref}[1]{\tablename~\ref{#1}}
\newcommand*{\figref}[1]{\figurename~\ref{#1}}
\newcommand*{\secref}[1]{Section~\ref{#1}}
\newcommand*{\appref}[1]{Appendix~\ref{#1}}
\renewcommand*{\eqref}[1]{equation~\ref{#1}}

\title{G~protein-coupled receptor-mediated calcium signaling in astrocytes}
\author{
        Maurizio De Pitt\`a\\
        EPI BEAGLE, INRIA Rh\^{o}ne-Alpes, Villeurbanne, France       
		\and
		Eshel Ben-Jacob\thanks{Deceased June 5, 2015.}\\
		School of Physics and Astronomy, Tel Aviv University, Ramat Aviv, Israel     
        \and
        Hugues Berry\\
        EPI BEAGLE, INRIA Rh\^{o}ne-Alpes, Villeurbanne, France       
        }

\begin{document}
\lstset{language=Python, basicstyle=\small\ttfamily,
              showstringspaces=false,
              columns=fixed}

\maketitle

\begin{abstract}
\noindent
Astrocytes express a large variety of G~protein-coupled receptors (GPCRs) which mediate the transduction of extracellular signals into intracellular calcium responses. This transduction is provided by a complex network of biochemical reactions which mobilizes a wealth of possible calcium-mobilizing second messenger molecules. Inositol 1,4,5-trisphosphate is probably the best known of these molecules whose enzymes for its production and degradation are nonetheless calcium-dependent. We present a biophysical modeling approach based on the assumption of Michaelis-Menten enzyme kinetics, to effectively describe GPCR-mediated astrocytic calcium signals. Our model is then used to study different mechanisms at play in stimulus encoding by shape and frequency of calcium oscillations in astrocytes.
\end{abstract}

\section{Introduction}
Calcium signaling is the most common measured readout of astrocyte activity in response to stimulation, be it by synaptic activity, by neuromodulators diffusing in the extracellular milieu, or by exogenous chemical, mechanical or optical stimuli. In this perspective, the individual astrocytic \ca\ transient is thought, to some extent, as an integration of the triggering stimulus \citep{PereaAraque_JNT2005}, and is thus regarded as an encoding or decoding of this stimulus, depending on the point of view \citep{Carmignoto2000,DePitta_FCN13}.

Multiple and varied are the spatiotemporal patterns of \ca\ elevations recorded from astrocytes in response to stimulation, each possibly carrying its own encoding \citep{Bindocci_Science2017}. Insofar as different encoding modes could correspond to different downstream signaling, including gliotransmission and thereby regulation of synaptic function, understanding the biophysical mechanisms underlying rich \ca\ dynamics in astrocytes is crucial. 

Calcium-induced \ca\ release (CICR) from the endoplasmic reticulum (ER) is arguably the best characterized mechanism of \ca\ signaling in astrocytes \citep{Zorec_ASN2012}. It ensues from nonlinear properties of \ca\ channels which are found on the ER membrane and are gated by the combined action of cytosolic \ca\ and the second messenger molecule inositol 1,4,5-trisphosphate (\ip3) \citep[][see also \textcolor{red}{Chapters~2--4}]{Shinohara_PNAS2011}. This second messenger molecule can be produced by the astrocyte either spontaneously or, notably, in response to activation by extracellular insults activation of G~protein-coupled receptors (GPCRs) found on the cell's plasma membrane \citep{ParriCrunelli_Neuroscience2003,Panatier_etal_Cell2011,Volterra_NRN2014}. Hence, \ip3\ together with these receptors, can be regarded as integral components of the interface whereby an astrocyte transduces extracellular insults into \ca\ responses \citep{Marinissen_TiPS2001}. Characterizing this interface is thus an essential step in our understanding of the emerging complexity of \ca\ signals, and we devote this chapter to this purpose. In the first part of the chapter, we will present a concise framework to model intracellular \ip3\ signaling in astrocytes. This framework is general and can easily be extended to include additional biological details, such as for example, the regulation of GPCR binding efficiency by protein kinase~C. Some of the models presented in this chapter are also subjected to revision and comparison with other astrocyte models in \textcolor{red}{Chapters~16} and~\textcolor{red}{18}.

\section{Modeling intracellular \ip3\ dynamics}\label{sec:ip3-modeling}
\subsection{Agonist-mediated \ip3\ production}\label{sec:PLCb-production}
G~protein-coupled receptors form a large family of receptors which owe their name to their extensively studied interaction with heterotrimeric~G proteins (composed of an $\upalpha$, $\upbeta$ and $\upgamma$ subunit) which undergo conformational changes that lead to the exchange of GDP for GTP, bound to the $\upalpha$-subunit, following receptor activation. Consequently, the G$\upalpha$- and G$\upbeta \upgamma$-subunits stimulate enzymes thereby activating or inhibiting the production of a variety of second messengers \citep{Marinissen_TiPS2001}.

Among all GPCRs, those that contain the G$\upalpha_\mathrm{q}$ subunit are linked with the cascade of chemical reactions that leads to \ip3\ synthesis. There, the G$\upalpha_\mathrm{q}$ subunit promotes activation of the enzyme pospholipase~C$\upbeta$ (\plcb) which hydrolizes the plasma membrane lipid phosphatidylinositol 4,5-bisphosphate (\pip2) into diacylglycerol (DAG) and \ip3\ \citep{RebecchiPentyala2000}. Examples of such receptors expressed by astrocytes ex vivo and in vivo are the type~I metabotropic glutamate receptor~1 and~5 (mGluR1/5) \citep{WangNedergaard2006,Sun_Science2013}, the purinergic receptor \ce{P2Y1} \citep{JourdainVolterra2007,DiCastro_Volterra_NatNeurosci2011,Sun_Science2013}, the muscarinic receptor mAchR1$\upalpha$ \citep{Takata_JN2011,Chen_PNAS2012,Navarrete_PB2012} and the adrenergic $\upalpha_1$ receptor \citep{Bekar_CC2008,Ding_CC2013}. While these receptors bind different agonists, and likely display receptor-specific binding kinetics, they all share the same downstream signaling pathway and therefore may be modeled in a similar fashion. 

Several are the available models for G$\upalpha_\mathrm{q}$-containing receptors, and the choice of what model to use rather than another depends on the level of biological detail and the questions one is interested in. Here our focus is on the rate of \ip3\ production upon activation of these receptors, so we wish to keep as simple as possible the description of the reactions that regulate the activation of \plcb\ by $\upalpha_\mathrm{q}$, $\upbeta$ and $\upgamma$ subunits. This is possible, assuming that these reactions are much faster than the downstream ones that result in \ip3\ production. In this case, a \textit{quasi steady-state approximation} (QSSA) holds whereby, in the series of reactions that leads from receptor agonist binding to activation of \plcb, the intermediate reactions involving the three receptor's subunits are at equilibrium on the time scale of the production of activated \plcb. Accordingly, assuming that on average the receptor at rest (\ce{R}) requires $n$ molecules of an agonist (\ce{A}) to promote activation of \plcb\ (\ce{R^*}) at rate $O_{N}$, we can write 
\begin{equation}
\ce{R + $n$A ->T[$O_{N}$] R$^*$}\label{rc:R-activation}
\end{equation}

We further make another assumption: that the cascade of reactions that leads to GPCR-mediated \ip3\ synthesis has a Michaelis-Menten kinetics (see \appref{app:Michaelis-Menten-kinetics}), so the \ip3\ production by~\plcb\ ($J_{\beta}$) can be taken proportional to the fraction of bound receptors,  defined as~$\Gamma_{A}=\ce{[R^{*}]/[R]_{T}}$, with~\ce{[R]_{T}=[R] + [R^{*}]} being the total receptor concentration at the site of \ip3\ production, i.e.,
\begin{equation}\label{eq:Jbeta}
J_{\beta} = O_\beta \cdot \Gamma_{A}
\end{equation}
In the above equation~$O_\beta$ is the maximal rate of~\ip3\ production by~\plcb\ and lumps information on receptor surface density as well as on the size of the \pip2 reservoir. Importantly, these two quantities may not be fixed, insofar as receptors are subjected to desensitization, internalization and recycling, and the reservoir of \pip2 could also be modulated by cytosolic~\ca\ and \ip3\ \citep{RheeBaeRev1997}. The reader interested in modeling these aspects may refer to \citet{Lemon_JTB2003}. In the following, we will assume $O_\beta$ constant for simplicity.

\begin{figure}[!tp]
\centering
	\begin{subfigure}[t]{0.48\textwidth}
		\caption{}
		\includegraphics[width=\textwidth]{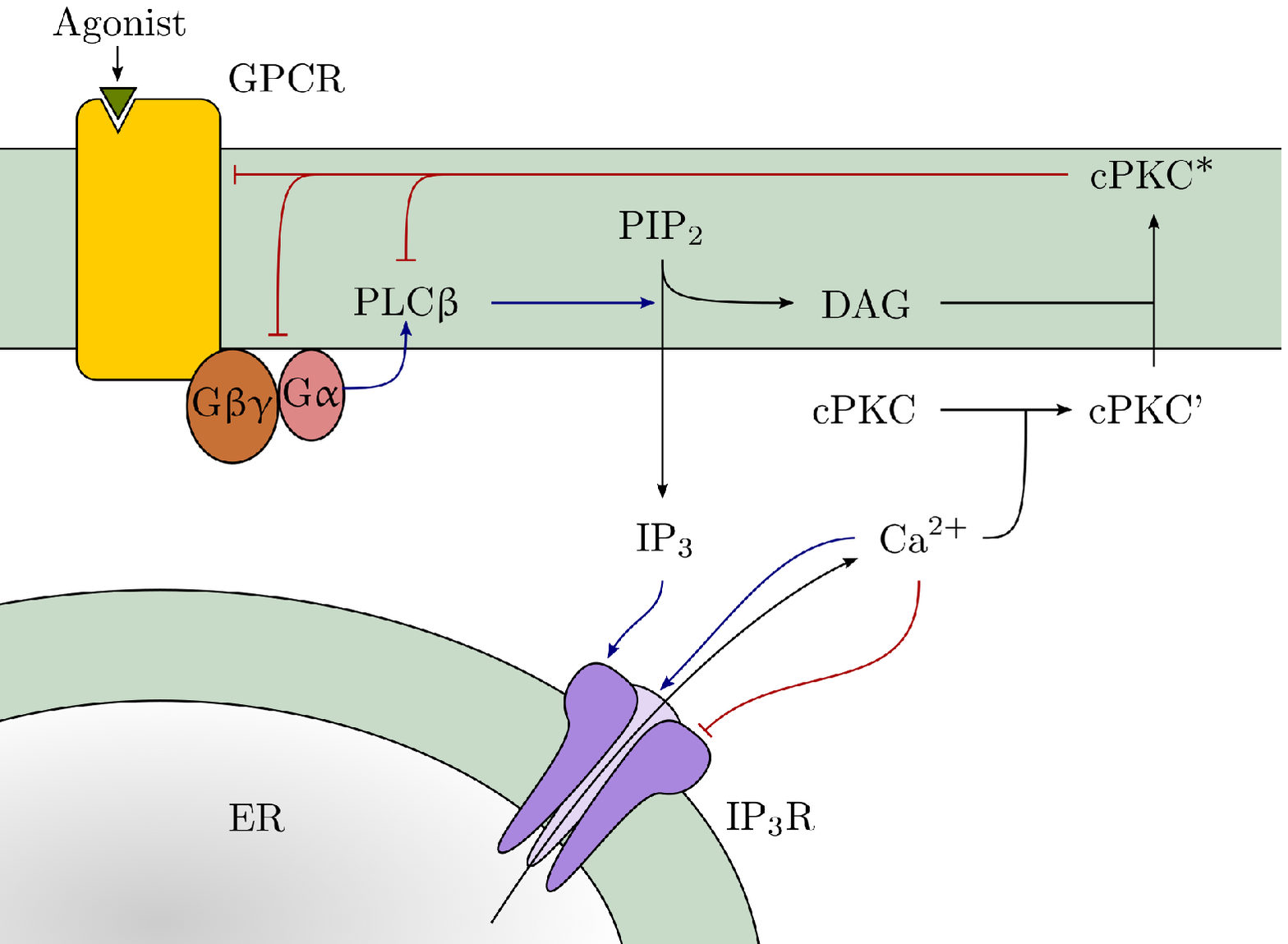}		
	\end{subfigure}\hfill
	\begin{subfigure}[t]{0.48\textwidth}
		\caption{}
		\includegraphics[width=\textwidth]{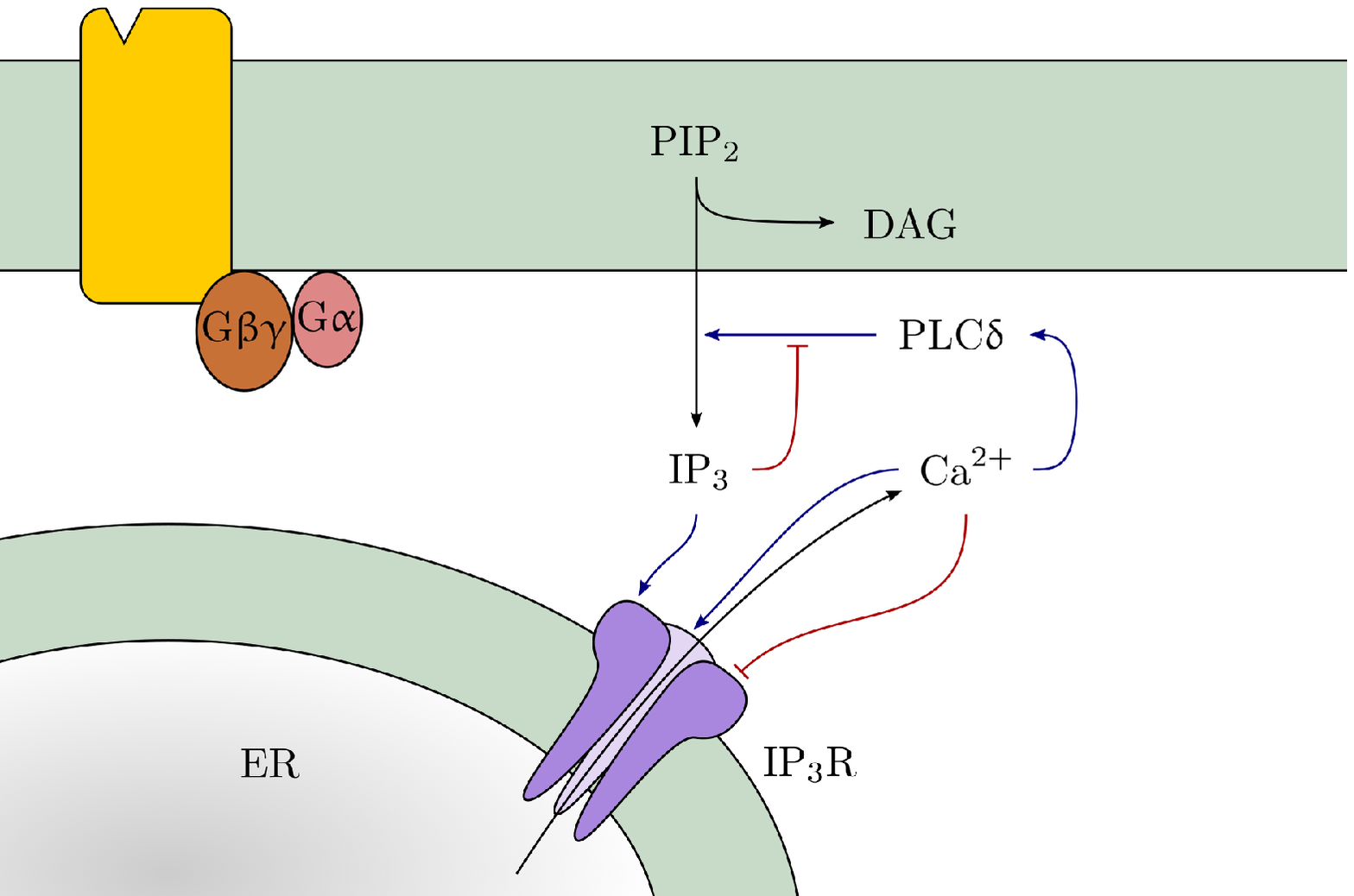}		
	\end{subfigure}
\caption{\ip3\ production.~\textbf{A}~Hydrolysis of the membrane lipid phosphatidylinositol 4,5-bisphosphate~(\pip2) by~\plcb\ and~\plcd\ isoenzymes produces~\ip3\ and diacylglycerol~(\dag). The contribution of~\plcb\ to~\ip3\ production depends on agonist binding to astrocyte G~protein-coupled receptors (GPCRs). This production pathway is inhibited via~receptor phosphorylation by \ca-dependent activation of conventional protein kinases~C~(cPKCs). Blue:~promoting pathway; \textit{red}:~inhibitory pathway.}
\label{fig:ip3-production}
\end{figure}

To seek an expression for~$J_{\beta}$, termination of~\plcb\ signaling has to be considered. With this regard, as illustrated in \figref{fig:ip3-production}A, there are two possible pathways whereby~\ip3\ production by \plcb\ ends \citep{RebecchiPentyala2000}. One is by reconstitution of the inactive G~protein heterotrimer, and coincides with unbinding of the agonist from the receptor, due to the intrinsic GTPase activity of the activated~G$\upalpha_\mathrm{q}$ subunit. The other is by phosphorylation of the receptor, the G$\upalpha_\mathrm{q}$~subunit,~\plcb\ or some combination thereof by conventional protein kinases~C (\ce{cPKC}) \citep{RyuRhee_JBC1990,CodazziTeruelMeyer2001}. This phosphorylation modulates either receptor affinity for agonist binding, or coupling of the bound receptor with the G~protein, or coupling of the activated G~protein with~\plcb, ultimately resulting in receptor desensitization \citep{Fisher_RevEJP1995}.

Denoting by~\ce{cPKC^*} the active, receptor-phosphorylating kinase~C, termination of~\plcb-mediated~\ip3\ production can then be modeled by the following pair of chemical reactions:
\begin{align}
\ce{R^*              & ->T[$\Omega_N$] R + $n$A           \label{rc:R-unbinding}\\
    cPKC^* + R^*     & <=>T[$O_{KR}$][$\Omega_{KR}$] cPKC^*-R^* ->T[$\Omega_{K}$] cPKC^* + R + $n$A}      \label{rc:R-PKC*-reaction}
\end{align}
From equations~\ref{rc:R-unbinding}--\ref{rc:R-PKC*-reaction} we have:
\begin{align}
\der{\ce{R^*}}{t}       &= O_N \ce{[A]^n} \ce{[R]} - \Omega_N \ce{[R^*]} - O_{KR} \ce{[cPKC^*][R^*]} + \Omega_{KR} \ce{[cPKC^*-R^*]} \label{eq:R*-ode}\\
\der{\ce{[cPKC^*-R^*]}}{t} &= O_{KR} \ce{[cPKC^*][R^*]} - (\Omega_{KR} + \Omega_K) \ce{[cPKC^*-R^*]} \label{eq:PR-ode}
\end{align}
Assuming that production of the intermediate kinase-receptor complex is at quasi steady state in reaction~\ref{rc:R-PKC*-reaction}, i.e. $\dd{\ce{[cPKC^*-R^*]}}/\dd{t} \approx 0$, provides (\eqref{eq:PR})
\begin{equation}
\ce{[cPKC^*-R^*]} = \frac{O_{KR}}{\Omega_{KR} + \Omega_K} \ce{[cPKC^*][R^*]} \label{eq:PR-qssa}
\end{equation}
Then, substituting this latter equation in \eqref{eq:R*-ode} gives
\begin{align}
\der{\ce{R^*}}{t} &= O_N \ce{[A]^n} \ce{[R]} - \Omega_N \ce{[R^*]} - O_{KR} \rpar{1-\frac{\Omega_{KR}}{\Omega_{KR} + \Omega_K}} \ce{[cPKC^*][R^*]} \nonumber \\
     &= O_N \ce{[A]^n} \ce{[R]} - \Omega_N \ce{[R^*]} - O_K \ce{[cPKC^*][R^*]} \label{eq:R*-ode-qssa}
\end{align}
where we defined $O_K = O_{KR} \rpar{1-\Omega_{KR}/\rpar{\Omega_{KR} + \Omega_K}}$.

To retrieve an equation for \ce{[cPKC^*]}, we consider the fact that activation of~\ce{cPKC} requires binding to the kinase of free cytosolic \ca\ ($C$) and~DAG, but only if~\ca\ binds first,~\ce{cPKC} can get sensibly activated by~DAG \citep{Oancea_Cell1998}. Accordingly, the following sequential binding reaction scheme for~\ce{cPKC} activation may be assumed:
\begin{align}
\ce{cPKC + Ca^{2+}} &\ce{<=>T[$O_{KC}$][$\Omega_{KC}$] cPKC$\,'$}\label{rc:PKC*-reaction-ca}\\
\ce{cPKC$\,'$ + DAG}&\ce{<=>T[$O_{KD}$][$\Omega_{KD}$] cPKC^*}\label{rc:PKC*-reaction-dag}
\end{align}
where~\ce{cPKC} is the inactive kinase, and~\ce{cPKC$\,'$} denotes the~\ca-bound kinase complex.
By~QSSA in reaction~\ref{rc:R-PKC*-reaction} it follows that the available activated kinase approximately equals to \ce{[cPKC^*]_T = [cPKC^*] + [cPKC^*-R^*] $\approx$ [cPKC^*]}. Moreover, it can be assumed that only a small fraction of~\ce{cPKC$\,'$} is bound by~DAG so that $\ce{[cPKC^*] \ll \ce{[cPKC$\,'$]}}$. In this fashion, the available~\ce{cPKC}, denoted by~\ce{[cPKC]_T}, can be approximated by~\ce{[PKC]_T$\,\approx\,$[PKC] + [PKC$\,'$]}. Accordingly, solving reactions~\ref{rc:PKC*-reaction-ca} and~\ref{rc:PKC*-reaction-dag} for \ce{[PKC$^{*}$]} provides
\begin{align}
\ce{[cPKC^*]} &= \rpar{\ce{[cPKC^*] + [cPKC$\,'$]}}\cdot \Hill{1}{\ce{[DAG]}}{K_{KD}}\nonumber\\
                  &\approx \ce{[cPKC$\,'$]} \cdot \Hill{1}{\ce{[DAG]}}{K_{KD}}\nonumber\\
                  &= \ce{[cPKC]_T}\cdot \Hill{1}{C}{K_{KC}} \cdot \Hill{1}{\ce{[DAG]}}{K_{KD}}\label{eq:PKC*-solution}
\end{align}
where $K_{KD}=\Omega_{KD}/O_{KD}$ and $K_{KC}=\Omega_{KC}/O_{KC}$, and $\Hill{1}{x}{K}$ denotes the Hill function $x/(x+K)$ (\appref{app:Hill-function}). In practice the activation of the kinase consists of two sequential translocations to the plasma membrane of its~C2 and~C1$_2$ domains \citep{Oancea_Cell1998}. The translocation of~C2 is regulated by \ca\, whereas that of~C1$_2$ is by~DAG. In this process however, experiments showed that the initial translocation of C2 is the rate limiting step for kinase activation \citep{ShinomuraNishizuka1991}, inasmuch as~C1$_2$ translocation rapidly follows that of~C2 \citep{CodazziTeruelMeyer2001}. This agrees with the notion that the \ce{cPKC} affinity for~DAG is regarded to be much higher than the affinity of the kinase for~\ca, i.e.~$K_{KD}\ll K_{KC}$ \citep{NishizukaRev1995}. Since the product of two Hill functions with widely separated constants can be approximated by the Hill function with the largest constant \citep{DePitta_JOBP2009}, \eqref{eq:PKC*-solution} can be rewritten as
\begin{equation}
\ce{[cPKC^*]} \approx \ce{[cPKC]_T} \cdot \Hill{1}{C}{K_{KC}}
\end{equation}
which, once replaced in \eqref{eq:R*-ode-qssa}, gives:
\begin{align}
\der{\ce{[R^*]}}{t} &= O_{N}\ce{[A]$^{n}$[R]} - \Omega_{N}\rpar{1+\frac{O_{K}\ce{[cPKC]_T}}{\Omega_{N}}\,\Hill{1}{C}{K_{KC}}}\ce{[R^*]} \label{eq:dRdt}
\end{align}
Finally, dividing both left and right terms in the above equation by \ce{[R]_T}, \eqref{eq:dRdt} can be rewritten as
\begin{equation}\label{eq:GammaA}
\der{\Gamma_{A}}{t} = O_{N}\ce{[A]$^{n}$}\,\rpar{1-\Gamma_{A}} - \Omega_{N}\rpar{1+\zeta \cdot \Hill{1}{C}{K_{KC}}}\,\Gamma_{A}
\end{equation}
where~$\zeta=O_{KC}\ce{[cPKC]_T}/\Omega_N$ quantifies the maximal receptor desensitization by~cPKC. In the approximation that receptor binding and activation is much faster than the effective~\plcb-mediated~\ip3\ production,~$\Gamma_{A}$ can be solved for the steady state. In this fashion,~\ip3\ production by~\plcb\ in~\eqref{eq:Jbeta} becomes
\begin{align}
J_{\beta} &= O_{\beta} \cdot \Hill{n}{\ce{[A]}}{\rpar{K_{N}\rpar{1+\zeta\,\Hill{1}{C}{K_{KC}}}}^\frac{1}{n}}\label{eq:Jbeta-ssa}
\end{align}
where~$K_{N}=\Omega_N/ O_N$. The Hill coefficient~$n$ denotes cooperativity of the binding reaction of the agonist with the receptor and is both receptor and agonist specific. For example, glutamate binding to subtype~1 mGluRs, such as those expressed by astrocytes \citep{GalloGhiani2000}, is characterized by negative cooperativity and found in association with a Hill coefficient of $n=0.48-0.88$ \citep{Suzuki2004}. On the contrary, binding of~ATP to P$_{2}$Y$_{1}$Rs of dorsal spinal cord astrocytes from rats is characterized instead by almost no cooperativity and $n=0.9-1$ \citep{Fam_JN2000}.

\subsection{\ip3\ production by receptors with $\upalpha$ subunits other than q-type}
A series of other astrocytic GPCRs, that traditionally associate with non-$\upalpha_\mathrm{q}$ subunits, have also been reported to mediate \ip3-triggered CICR, both in situ and in vivo. These include G$\upalpha_{\mathrm{i/o}}$-coupled \ce{GABA_B} receptors \citep{KangNedergaard1998,Serrano_etal_JN2006,Mariotti_Glia2016}, endocannabinoid \ce{CB_1} receptors \citep{NavarreteAraque_Neuron2008,MinNevian_NatNeurosci2012},  adenosinergic \ce{A_1} receptors \citep{Cristovao-Ferreira_PS2013}, adrenergic $\upalpha_2$ receptors \citep{Bekar_CC2008}, and dopaminergic \ce{D_{2/3}} receptors \citep{Jennings_Glia2017}; as well as G$\upalpha_{\mathrm{s}}$-coupled receptors like adenosine \ce{A_{2A}} receptors \citep{Cristovao-Ferreira_PS2013}, and dopamine \ce{D_{1/5}} receptors \citep{Jennings_Glia2017}. $\upalpha_\mathrm{i/o}$ and $\upalpha_\mathrm{s}$ subunits are not expected to be linked with \ip3\ synthesis \citep{Marinissen_TiPS2001}, rather they respectively inhibit or stimulate intracellular production of cAMP. Therefore the mechanism whereby these receptors could also promote mobilization of \ca\ from \ip3-sensitive ER stores remains a matter of investigation.

One obvious possibility is that some of these receptors could be atypical in astrocytes and also be coupled with G$\upalpha_\mathrm{q}$, as it seems the case for example of astrocytic \ce{CB1Rs} in the hippocampus \citep{NavarreteAraque_Neuron2008} and in the basal ganglia \citep{Martin_Science2015}. Biased agonism could also be another possibility since the spatiotemporal pattern of agonist action on GPCRs could be quite different depending on agonist-binding kinetics of the receptor, especially if agonists differentially engage dynamic signalling and regulatory processes \citep{Overington_NRDD2006}, such as in the likely scenario of synapse-astrocyte interactions \citep{Heller_Glia2015}. However, there is not yet direct structural evidence for distinct receptor conformations linked to specific signals such as distinct G~protein classes, and future studies are required to compare crystal structures of astrocytic GPCRs bound to biased and unbiased ligands to establish these relationships \citep{Violin_TiPS2014}.

Alternatively, other signaling pathways mediated by cAMP that result in CICR could also be envisaged. In particular, \citet{Doengi_PNAS2009} reported that GABA-evoked astrocytic \ca\ events in the olfactory bulb are fully prevented by blockers of astrocytic GABA transporters (GATs), but only partially by \ce{GABA_B} antagonists. GAT activation leads to an increase of intracellular \ce{Na^+}, since this ion is cotransported with GABA, and such increase indirectly inhibits the \ce{Na^+/Ca^{2+}} exchanger on the plasma membrane. In turn, the ensuing \ca\ increase could be sufficient to induce \ca\ release from internal stores by stimulation of endogenous \ip3\ production \citep[][see the following Section]{Losi_PTRSB2014}. This possibility is further corroborated by the observation that astrocytic GATs could indeed be inhibited or stimulated respectively by \ce{A1Rs} or \ce{A_{2A}Rs} \citep{Cristovao-Ferreira_PS2013}. 

Yet other mechanisms could be at play for different receptors. Dopaminergic receptors for example could either increase (\ce{D_{1/5}} receptors) or decrease (\ce{D_{2/3}} receptors) intracellular \ca\ levels in astrocytes \citep{Jennings_Glia2017}. This could indeed be explained assuming a possible action of these receptors on GATs  which, similarly to adenosinergic receptors, could respectively increase or reduce \ce{GABA/Na^+} cotransport into the cell, ultimately promoting or inhibiting CICR according to what was suggested for \ce{GABA_BRs}. However there is also evidence that nontoxic levels of dopamine could be metabolized by monoamine-oxidase in cultured astrocytes, resulting in the production of hydrogen peroxide \citep{Vaarmann_JBC2010}. This reactive oxygen species ultimately activates lipid peroxidation in the neighboring membranes which in turn triggers PLC-mediated \ip3\ production and CICR. Overall these different scenarios unravel additional complexity in the possible mechanisms of GPCR-mediated CICR in astrocytes and call for future modeling efforts that are beyond the scope of this chapter.

\subsection{Endogenous \ip3\ production}
Phospholipase~C$\updelta$~(\plcd) is the enzyme responsible of endogenous \ip3\ production in astrocytes, that is~\ip3\ production that does not require external (i.e.~exogenous) stimulation \citep{Ochocka_ABP2003,Suh_BMB2008}. The specific catalytic activity of this enzyme in the presence of cytosolic \ca\ is 50- to 100-fold greater than \ca-stimulated activity of \plcb\ in the absence of activating G protein subunits \citep{RebecchiPentyala2000}, suggesting that \plcd\ is prominently activated by increases of intracellular \ca\ \citep{RheeBaeRev1997}.

\figref{fig:ip3-production}B exemplifies the biochemical network associated with~\plcd\ activation. Structural and mutational studies of~\plcd\ complexes with~\ca\ and~\ip3, revealed complex interactions of~\ca\ with several negatively charged residues within the \plcd\ catalytic domain \citep{EssenWilliams_Nature1996,EssenWilliams_Biochem1997,RheeBaeRev1997}, hinting cooperative binding of at least two \ca\ ions with this enzyme \citep{EssenWilliams_Biochem1997}. In agreement with these experimental findings, we model~\plcd-mediated \ip3\ production~($J_{\delta}$) as \citep{PawelczykMatecki_EurJBiochem1997,HoferGiaume2002}:
\begin{equation}\label{eq:Jdelta}
J_{\delta} = \hat{J}_{\delta}(I)\cdot\Hill{2}{C}{K_{\delta}}
\end{equation}
where \Hill{2}{C}{K_{\delta}} denotes the Hill function of~$C$ with coefficient~2 and affinity~$K_{\delta}$ (\appref{app:Parameters}), and~$\hat{J}_{\delta}(I)$ is the maximal rate of~\ip3\ production by~\plcd\ which depends on intracellular~\ip3\ ($I$). Experiments revealed that high~\ip3\ concentrations, i.e.~$>\SI{1}{\micro \Molar}$, inhibit~\plcd\ activity by competing with \pip2\ binding to the enzyme \citep{AllenBarres_Nature2009}. Accordingly, the maximal~\plcd\-dependent~\ip3\ production rate can be modeled by
\begin{equation}\label{eq:Od-max}
\hat{J}_{\delta}(I)=\frac{O_{\delta}}{1+\frac{I}{\kappa_{\delta}}}=O_{\delta}\rpar{1-\Hill{1}{I}{\kappa_\delta}}
\end{equation}
where~$O_{\delta}$ is the maximal rate of~\ip3\ production by~\plcd\ and~$\kappa_{\delta}$ is the inhibition constant of~\plcd\ activity.

\subsection{\ip3\ degradation}
\begin{figure}[!tp]
\centering
	\begin{subfigure}[t]{0.7\textwidth}
		\caption{}
		\includegraphics[width=\textwidth]{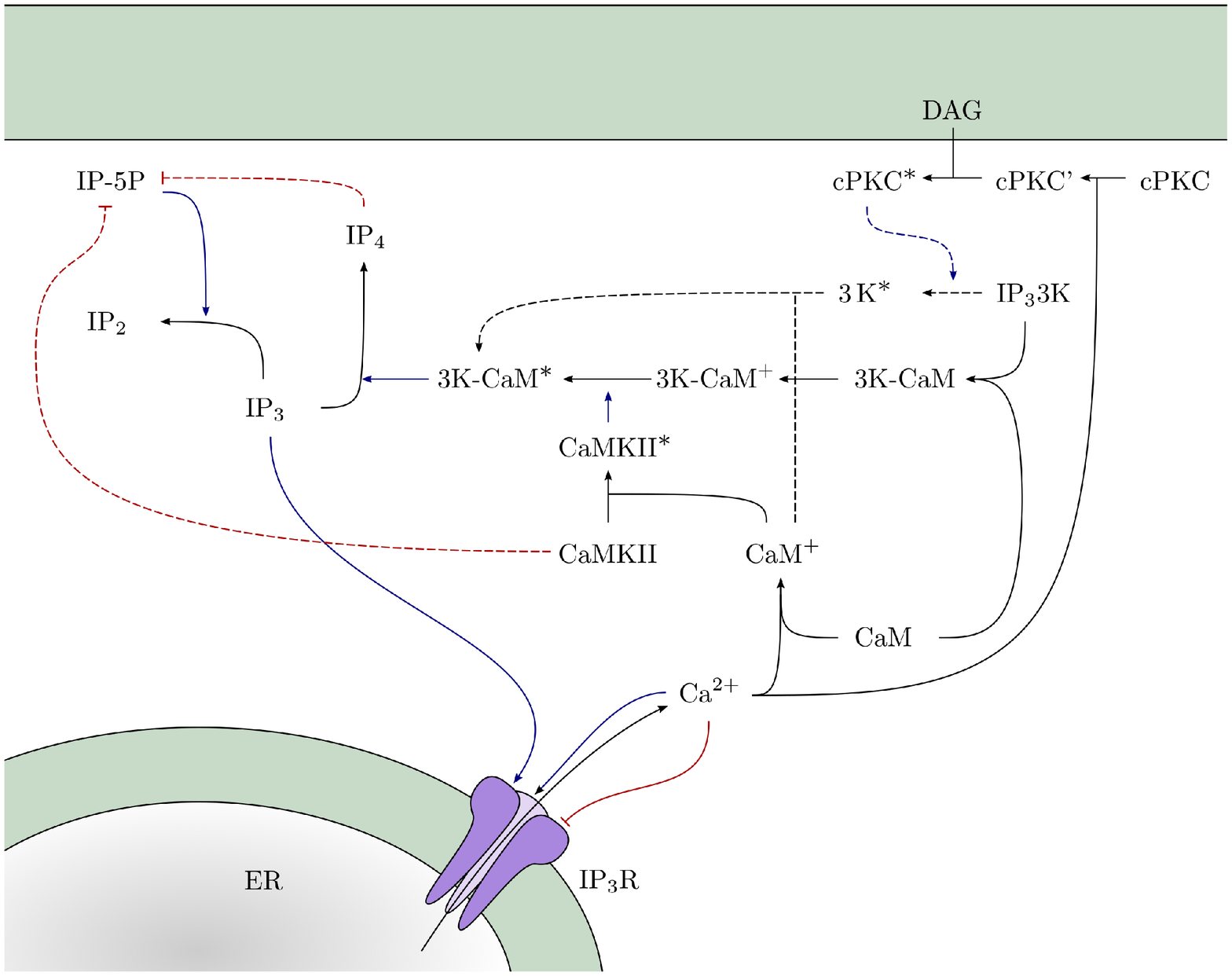}
	\end{subfigure}\\[1ex]
	\begin{subfigure}[t]{0.7\textwidth}
		\caption{}
		\includegraphics[width=\textwidth]{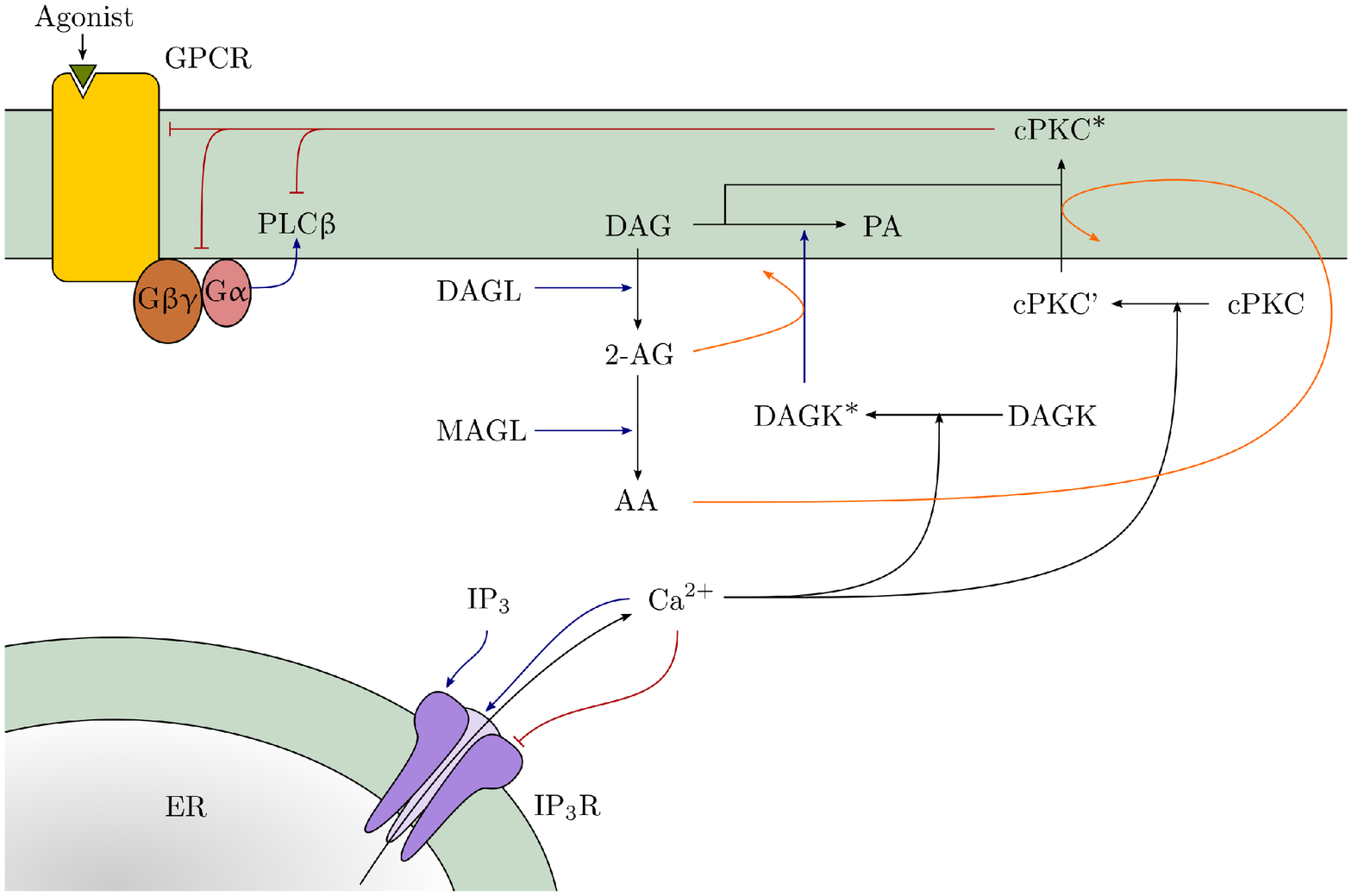}		
	\end{subfigure}
\caption{\ip3\ and \dag\ degradation.~\textbf{A}~Degradation of~\ip3\ occurs by phosphorylation into inositol 1,3,4,5-tetrakisphosphate~(\ce{IP4}) by~\3k\, and dephosphorylation into lower inositol phosphates by~\5p. Both pathways are regulated by~\ca: \3k\ activity is stimulated by phosphorylation by~\ca /calmodulin-dependent protein kinase~II~(CaMKII), whereas~\5p\ is inhibited thereby. Moreover~\3k-mediated degradation could also be promoted by~\ca\ and \dag-dependent~cPKC-mediated phosphorylation, while \5p\ could also be inhibited by \ce{IP4}. For the sake of simplicity, \5p\ dependence on \ca\ and \ce{IP4} along with \3k\ dependence on~cPKC are not taken into consideration in this study (\textit{dashed pathways}). \textbf{B}~\dag\ is mainly degraded into phosphatidic acid~(PA) by~DAG kinases~(DAGK) in a \ca-dependent fashion, and to a minor extent, into 2-arachidonoylglycerol (2-AG) by DAG lipases~(DAGL). In turn~2-AG is hydrolized by monoacylglycerol lipase~(MAGL) into arachidonic acid~(AA). 2-AG~and~AA may promote activity of DAGK and \cpkc\ (\textit{orange patwhays}) although this scenario is not taken into consideration here. Colors of other pathways as in \figref{fig:ip3-production}.}
\label{fig:ip3-degradation}
\end{figure}

There are two pathways for~\ip3\ degradation in astrocytes. The first one is by dephosphorylation of~\ip3\ by inositol polyphosphate 5-phosphatase~(\5p). The other one occurs through phosphorylation of~\ip3\ by the~\ip3\ 3-kinase~(\3k). Both pathways could be~\ca\ dependent but in opposite ways: while the activity of~\3k\ is stimulated by cytosolic~\ca\ \citep{CommuniErneux1997}, \5p\ is inhibited instead \citep{CommuniErneux2001} (\figref{fig:ip3-degradation}A). Thus, depending on the~\ca\ concentration in the cytoplasm, different mechanisms of~\ip3\ degradation could exist \citep{SimsAllbritton1998}. Moreover,~\5p-mediated~\ip3\ degradation could also be inhibited by competitive binding of inositol 1,3,4,5-tetrakisphosphate~(IP$_{4}$) produced by~\3k-mediated~\ip3\ phosphorylation \citep{Connolly_JBC1987,Erneux_BBA1998}, thereby making the two degradation pathways interdependent \citep{Hermosura_Nature2000}. However, we will not consider this aspect any further, since modeling of this reaction pathway requires a detailed consideration of the complex metabolic network underpinning degradation of the large family of inositol phosphates \citep{CommuniErneux2001,Irvine_NRMCB2001}. The reader interested in these aspects may refer to \citet{DupontErneux1997} for a sample modeling approach to the problem.

Both~\5p-mediated dephosphorylation~($J_{5P}$) and~\3k-mediated phosphorylation of~\ip3 ($J_{3K}$) can be described by Michaelis-Menten kinetics \citep{IrvineBerridge_Nat1986,TogashiOnaya_BiochemJ1997}, i.e.,
\begin{align}
J_{5P} &= \hat{J}_{5P}\cdot \Hill{1}{I}{K_{5}}\label{eq:J5P-generic}\\
J_{3K} &= \hat{J}_{3K}(C)\cdot \Hill{1}{I}{K_{3}}\label{eq:J3K-generic}
\end{align}
Since~$K_{5P}>\SI{10}{\micro \Molar}$ \citep{VerjansErneux1992,SimsAllbritton1998}, and such high~\ip3\ concentrations are unlikely to be physiological \citep{Lemon_JTB2003,KangOthmer_Chaos2009}, the activity of~\5p\ can be assumed far from saturation. Accordingly, the~\ip3\ degradation rate by~\5p\ can be linearly approximated by \citep{StryerBiochemistryBOOK}:
\begin{equation}\label{eq:J5P-linear-approx}
J_{5P} \approx \Omega_{5P}\cdot I
\end{equation}
where~$\Omega_{5P}=\hat{J}_{5P}/K_{5}$ is the maximal rate of~\5p-mediated \ip3\ degradation in the linear approximation.

\ip3\ phosphorylation by~\3k\ is regulated in a complex fashion (\figref{fig:ip3-degradation}A). For resting conditions, when intracellular~\ip3\ and~\ca\ concentrations are below~0.1~$\mu$M, \citep{ParpuraHaydon2000,MishraBhalla2002,KangOthmer_Chaos2009}, it is very slow. On the other hand, as \ca\ increases, \3k\ activity is substantially stimulated by~its phosphorylation by~CaMKII in a \ca/calmodulin~(CaM)--dependent fashion \citep{CommuniErneux1997}. A further possibility could eventually be that \3k\ is also inhibited by~\ca-dependent PKC~phosphorylation \citep{SimRhee_JBC1990}, however, since evidence for the existence of such inhibitory pathway is contradictory \citep{CommuniErneux_Rev1995}, this possibility will not be taken into further consideration in this study.

Phosphorylation of~\3k\ by active~\ce{CaMKII} (i.e.~\ce{CaMKII^*}) only occurs at a single threonine residue \citep{CommuniErneux1997,CommuniErneux_JBC1999}, so that it can be assumed that the rate of \3k\ phosphorylation is $J_{3K}^*(C)\propto \ce{[CaMKII^*]}$. On the other hand, activation of~\ce{CaMKII} is~\ca/CaM-dependent and occurs in a complex fashion because of the unique structure of this kinase, which is composed of~$\sim$12 subunits, with three to four phosphorylation sites each \citep{KolodziejStoops_JBC2000}. Briefly,~\ca\ increases lead to the formation of a \ce{Ca^{2+}-CaM} complex~(\ce{CaM^+}) that may induce phosphorylation of some of the sites of each~\ce{CaMKII} subunit. However, only when two of these sites at neighboring subunits are phosphorylated, CaMKII~quickly and fully activates \citep{HansonSchulman_Neuron1994}. Despite the multiple~\ce{CaM^+} binding reactions in the inactive kinase, experiments showed that~\ce{KII} activation by~\ce{CaM^+} can be approximated by a Hill equation with unitary coefficient \citep{DeKonickSchulman1998}. Hence, the following kinetic reaction scheme for~\ce{CaMKII} phosphorylation can be assumed:
\begin{align}
\ce{4 Ca^{2+} + CaM} & \ce{<=>T[$O_{0}$][$\Omega_{0}$] CaM^+}\label{rc:Ca-binding}\\
\ce{KII + CaM^+}   & \ce{<=>T[$O_{b}$][$\Omega_{b}$] CaMKII <=>T[$\Omega_{a}$][$\Omega_{i}$] CaMKII^*} \label{rc:KII-activation}
\end{align}
Consider then first the binding reaction in~\ref{rc:KII-activation}. Assuming that the second step is very rapid with respect to the first one \citep{ThielGreengard_PNAS1988,DeKonickSchulman1998}, the generation of~\ce{CaMKII^*} is in equilibrium with~\ce{CaMKII} consumption, i.e.,
\begin{align}\label{eq:CaMKII*}
\ce{[CaMKII^*]} &\approx \frac{\Omega_{a}}{\Omega_{i}}\ce{[CaMKII]}
\end{align}
Then, under the hypothesis of quasi-steady state for~CaMKII,
\begin{align}
\der{\ce{[CaMKII]}}{t} &= O_{b}\,\ce{[KII][CaM^+]}-\rpar{\Omega_{a}+\Omega_{b}}\ce{[CaMKII]}+\Omega_{i}\,\ce{[CaMKII^*]}\approx 0
\end{align}
Replacing \ce{[CaMKII$^{*}$]} from~\eqref{eq:CaMKII*} in the latter equation provides
\begin{align}\label{eq:CaMKII*-equation-2}
\ce{[CaMKII^*]} &= K_{a}K_{b}\ce{[KII][CaM^+]}
\end{align}
where~$K_{a}=\Omega_{a}/ \Omega_{i}$ and~$K_{b}= O_{b}/ \Omega_{b}$. Defining the total kinase~II concentration as\linebreak[4]\ce{[KII]_T = [KII] + [CaMKII] + [CaMKII^*]} and assuming it constant, \eqref{eq:CaMKII*-equation-2} can be rewritten as
\begin{align}\label{eq:CaMKII*-equation-3}
\ce{[CaMKII$^{*}$]} &= \frac{K_{a}\ce{[KII]$_{T}$}}{1+K_{a}}\cdot \Hill{1}{\ce{[CaM$^{+}$]}}{K_{m}}
\end{align}
with~$K_{m}=\rpar{K_{b}\rpar{1+K_{a}}}^{-1}$.

The substrate concentration for the enzyme-catalyzed reaction~\ref{rc:KII-activation} is provided by reaction~\ref{rc:Ca-binding} and reads (by QSSA)
\begin{align}\label{eq:CaM+}
\ce{[CaM$^{+}$]} &= \ce{[CaM]}\cdot \Hill{4}{C}{K_0}
\end{align}
with~$K_{0}=O_{0}/ \Omega_{0}$. Therefore, replacing the latter expression for \ce{[CaM^+]} in~\eqref{eq:CaMKII*-equation-3}, finally provides
\begin{align}
\ce{[CaMKII^*]} &= \frac{K_{a}\ce{[KII]_T}}{1+K_a}\,\rpar{1+\frac{K_m}{\ce{[CaM]}}}^{-1}\cdot \Hill{4}{C}{\frac{K_0 K_m}{K_m + \ce{[CaM]}}}
\end{align}
Defining the~\ca\ affinity constant of~\3k\ as~$K_{D}=K_0 K_m /\rpar{K_m +\ce{[CaM]}}$, the above calculations show that, despite its complexity, the reaction cascade underlying the activation of~\ce{CaMKII} can be concisely described by a Hill function of the~\ca\ concentration~($C$) so that~$\ce{[CaMKII^*]}\propto \Hill{4}{C}{K_D}$. Accordingly, it is also $\hat{J}_{3K}(C)\propto \Hill{4}{C}{K_D}$, and \eqref{eq:J3K-generic} for \3k-mediated \ip3\ degradation can be rewritten as
\begin{equation}\label{eq:J3K-exact}
J_{3K} = O_{3K}\cdot \Hill{4}{C}{K_{D}} \Hill{1}{I}{K_{3}}
\end{equation}
where~$O_{3K}$ is the maximal rate of~\ip3\ degradation by~\3k.

\section{Encoding of stimulation by combined \ip3\ and \ca\ dynamics}
\subsection{The \gchi\ model for \ip3/\ca\ signaling}
A corollary of the biological and modeling arguments exposed in the previous section is that \ca\ and \ip3\ signals are, generally speaking, dynamically coupled in astrocytes. This implies that a complete model that mimics astrocytic \ip3\ signaling must also include a description of CICR. An example of such models is the so-called $ChI$ model originally introduced by \citet{DePitta_JOBP2009}, which is constituted by three ODEs respectively for intracellular \ca\ ($C$), the \ip3R gating variable $h$ and the mass-balance equation for intracellular \ip3\ lumping terms, (\ref{eq:Jdelta}), (\ref{eq:J5P-linear-approx}) and~(\ref{eq:J3K-exact}), i.e.
\begin{align}
	\der{C}{t} &= J_r(C,h,I) + J_l(C) - J_p(C) \label{eq:C}\\
	\der{h}{t} &= \Omega_h(C,I)\rpar{h_\infty(C,I)-h} \label{eq:h}\\
	\der{I}{t} &= O_{\delta}\Hill{2}{C}{K_\delta}\rpar{1-\Hill{1}{I}{\kappa_\delta}} -O_{3K}\, \Hill{4}{C}{K_{D}} \Hill{1}{I}{K_{3}} -\Omega_{5P}\, I \label{eq:I}
\end{align}
The above model can be extended to explicitly modeling of GPCR dynamics by a \gchi\ model. To this aim, we add to the right-hand side of \eqref{eq:I} the contribution of GPCR-mediated \ip3\ synthesis given by \eqref{eq:Jbeta-ssa}. However, if one is interested in how GPCR kinetics evolves with \ip3\ and \ca\ dynamics, then the formula for $J_\beta$ given by \eqref{eq:Jbeta} must be used instead of  \eqref{eq:Jbeta-ssa}. Accordingly, the above system of equations must be completed by \eqref{eq:GammaA} for astrocytic receptor activation, i.e. 
\begin{align}
\der{\Gamma_A}{t} &=\ldots \tag{\ref{eq:GammaA}}\\
\der{C}{t} &= \ldots \tag{\ref{eq:C}}\\
\der{h}{t} &= \ldots \tag{\ref{eq:h}}\\
\der{I}{t} &= O_\beta\Gamma_A + O_{\delta}\Hill{2}{C}{K_\delta}\rpar{1-\Hill{1}{I}{\kappa_\delta}} -O_{3K}\, \Hill{4}{C}{K_{D}} \Hill{1}{I}{K_{3}} -\Omega_{5P}\, I \label{eq:gchi-I}
\end{align}
Regarding the differential equations for the variables $C$ and $h$ above, the original formulation of the \gchi\ model considered the Li-Rinzel description for CICR previously introduced in \textcolor{red}{Chapter~3} \citep{LiRinzel1994}. In the following, we will refer to this formulation. In practice however, it must be noted that any suitable model of \ca\ and \ip3R dynamics discussed in \textcolor{red}{Chapters~2,~3} and~\textcolor{red}{16} can be adopted in lieu of the Li-Rinzel description, and accordingly different models of \gchi\ type may be developed, each possibly customized to study specific aspects of coupled \ip3\ and \ca\ signaling in astrocytes.

\begin{figure}[!tp]
	\centering
	\includegraphics[width=0.9\textwidth]{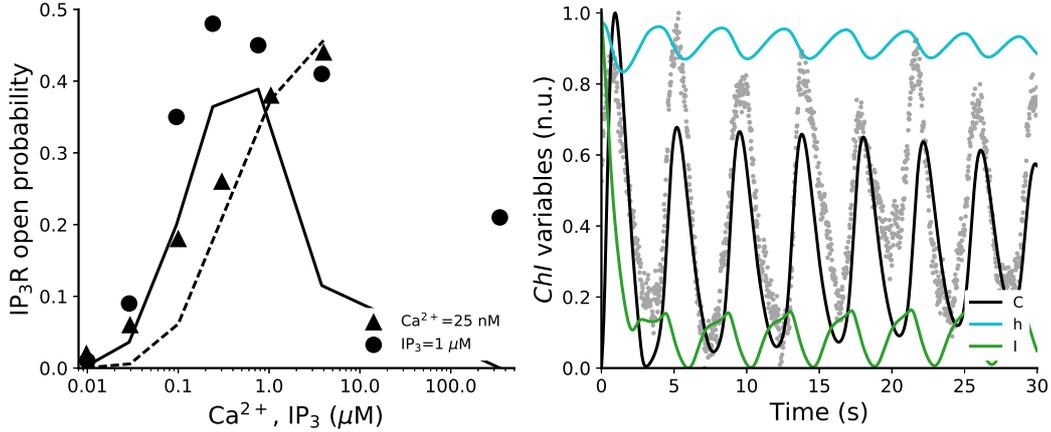}
	\caption{\textbf{\gchi\ model}. (\textit{left panel})~Fit of \ip3Rs kinetic parameters on experimental data of steady-state open probabilities of type-2 \ip3Rs by \citet{Ramos_BJ2000}. In this example, and through all this chapter, we consider the Li-Rinzel description for CICR. This choice allows a reasonable fit (\textit{solid and dashed lines}) of the receptors' open probability as function of either intracellular \ip3\ ($\blacktriangle$) or  intracellular \ca\ ($\CIRCLE$). The only exception is for \ca\ concentrations $>\SI{1}{\micro \Molar}$ for which the open probability predicted by the Li-Rinzel model (\textit{solid line}) vanishes much more quickly than experimental values. (\textit{right panel}) Sample \ca\ ($C$), \ip3\ ($I$) and $h$ traces ensuing from a simulation of the \gchi\ model to reproduce experimental \ca\ oscillations in cultured astrocytes (\textit{gray data points}) triggered by application of $>\SI{5}{\micro \Molar}$ glutamate. Experimental data courtesy of Nitzan Herzog (University of Nottingham). A saturating glutamate concentration (i.e. $\Gamma_A=1$) was assumed with initial conditions $C(0) = \SI{0.098}{\micro \Molar},\,h(0)=0.972$ and $I(0)=\SI{0.190}{\micro \Molar}$. Simulated \ca\ and \ip3\ traces are reported in normalized units with respect to minimum values of $C_0=\SI{0.1}{\micro \Molar}$ and $I_0=\SI{0.16}{\micro \Molar}$ and peak values of $\hat{C}=\SI{1.42}{\micro \Molar}$ and $\hat{I}=\SI{0.19}{\micro \Molar}$. Model parameters as in \tabref{tab:Model-Parameters} except for $O_\beta=\SI{0.141}{\micro \Molar s^{-1}}$ and $O_{3K}=\SI{0.163}{\micro \Molar s^{-1}}$.}
	\label{fig:chi}
\end{figure}

\figref{fig:chi} illustrates some characteristics of \ip3\ and \ca\ dynamics reproduced by the \gchi\ model. In the left panel of this figure, \ip3R kinetic parameters are chosen to fit, as closely as possible, experimental data points for the steady-state open probabilities of type-2 \ip3Rs at fixed \ca\ (\textit{solid line}) and \ip3\ concentrations (\textit{dashed line}). In the right panel, the remainder of the parameters of the model are then set to reproduce (\textit{solid black line}) a sample \ca\ trace imaged by confocal microscopy on cultured astrocytes (\textit{gray data points}). It may be observed how the associated \ip3\ and $h$ oscillations predicted by the model, are almost out of phase with respect to the \ca\ ones. For $h$, this is due to \ip3R kinetics, whereby an increase of cytosolic \ca\ promotes receptor inactivation. For \ip3\ instead, this dynamics is a direct consequence of the \ca-dependent rate of degradation of this molecule by the \3k\ enzyme. This is a crucial aspect of intracellular \ip3\ regulation in astrocytes which is addressed more in detail below.

\subsection{Different regimes of \ip3\ signaling}
\begin{figure}[!tp]
	\centering
	\includegraphics[width=0.9\textwidth]{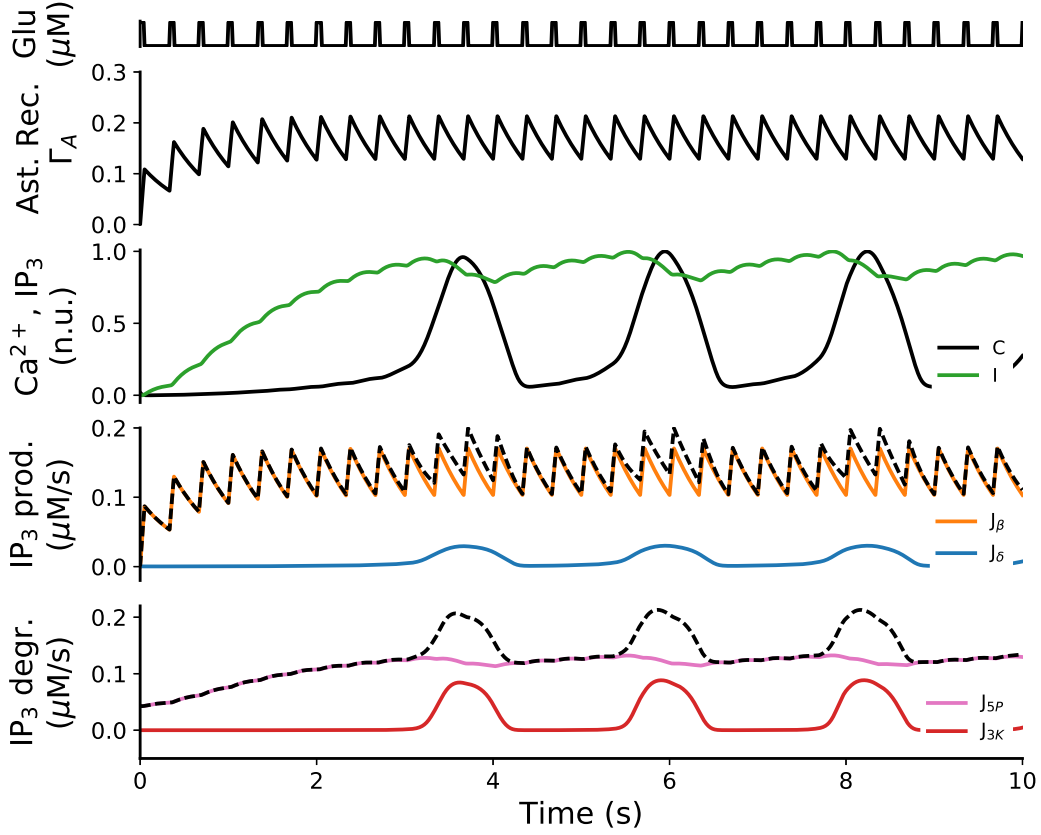}
	\caption{Coexistence of different regimes of \ip3\ signaling. From top to bottom: (\textit{first panel})~Repetitive stimulation of an astrocyte by puffs of glutamate ($\SI{8}{\micro \Molar}$, rectangular pulses at rate \SI{0.33}{Hz} and 15\% duty cycle); (\textit{second panel})~fraction of activated astrocytic receptors; (\textit{third panel})~ensuing \ca\ ($C$) and \ip3\ ($I$) traces (normalized with respect to their maximum excursion: $C_0=\SI{40}{n \Molar},\,I_0=\SI{50}{n \Molar},\,\hat{C}=\SI{0.73}{\micro \Molar},\,\hat{I}=\SI{0.15}{\micro \Molar}$); (\textit{fourth panel})~total rate of \ip3\ production (\textit{dashed line}) and contributions to it by \plcb\ ($J_\beta$) and \plcd\ ($J_\delta$); (\textit{bottom panel})~total rate of \ip3\ degradation (\textit{dashed line}) resulting from the combination of degradation by \5p\ ($J_{5P}$) and \3k\ ($J_{3K}$). Besides \ca\ pulsed-oscillations, \ip3\ is mainly regulated by \plcb\ (\textit{orange trace}) and \5p (\textit{violet trace}), and its concentration tends to increase in an integrative fashion with the number of glutamate puffs. During \ca\ elevations instead, activity of \plcd\ (\textit{blue trace}) and \3k\ (\textit{red trace}) become significant, with this latter responsible for a sharp drop of intracellular \ip3. Model parameters as in \tabref{tab:Model-Parameters} except for $C_T=\SI{10}{\micro \Molar},\,O_P=\SI{10}{\micro \Molar s^{-1}}$ and $O_\delta=\SI{0.05}{\micro \Molar s^{-1}}$.}
	\label{fig:gchi-dynamics}
\end{figure}

To develop the \gchi\ model in \secref{sec:ip3-modeling}, we stressed on the molecular details of the \ca\ dependence of the different enzymes involved in \ip3\ signaling, yet how this dependence shapes \ca\ and \ip3\ oscillations remains to be elucidated. With this purpose, we consider in \figref{fig:gchi-dynamics} the simple scenario of \ca\ oscillations triggered by repetitive stimulation of an astrocyte by puffs of extracellular glutamate (\textit{top three panels}), and look at the different contributions to \ip3\ production and degradation underpinning the ensuing \ca\ and \ip3\ dynamics (\textit{lower panels}). With this regard, it may be noted how the total rate of \ip3\ production (\textit{dashed line} in the \textit{fourth panel} from top) almost resembles the dynamics of activation of astrocyte receptors ($\Gamma_A$, \textit{second panel} from top) except for little bumps in correspondence of \ca\ pulse-like elevations (\textit{solid trace}, \textit{third panel} from top). Consideration of the different contributions to \ip3\ by \plcb\ (\textit{orange trace}) and \plcd\ (\textit{blue trace}) reveals that, while most of \ip3\ production is driven by mGluR-mediated \plcb\ activation, those bumps are instead caused by \plcd, whose activation is substantially boosted during intracellular \ca\ elevations.

Similar arguments also hold for \ip3\ degradation (\textit{bottom panel}). In this case, the total rate of \ip3\ degradation (\textit{dashed line}) closely mimics \ip3\ dynamics in between \ca\ elevations (\textit{green trace}, \textit{third panel} from top), and is mostly contributed by \ca-independent \5p-mediated degradation (\textit{violet trace}). This scenario however changes during \ca\ elevations, when \3k\ activation becomes significant and promotes faster rates of \ip3\ degradation, as mirrored by the \textit{dashed line} which peaks in correspondence of \ca\ oscillations.

Overall, these observations suggest that \ca-independent activity of \plcb\ and \5p\ vs. \ca-dependent activation of \plcd\ and \3k\ account for different regimes of \ip3\ signaling. One regime corresponds to low intracellular \ca\, close to resting concentrations, whereby \ip3\ is mainly produced by receptor-mediated activation of \plcb\ against degradation by \5p. The other regime significantly adds to the former for sufficiently high \ca\ elevations, where \ip3\ production is boosted by \plcd, but also \ip3\ degradation is faster by \3k\ activation.

The contribution to \ip3\ production and degradation by each enzyme clearly depends on their intracellular expression as reflected by the values of the rate constants $O_\beta,\, O_\delta,\,O_{3K}$ and $\Omega_{5P}$ in \eqref{eq:gchi-I}. Nonetheless, it should be noted that the existence of different regimes of \ip3\ production and degradation is regardless of these rate values, insofar as it is set by the values of the Michaelis-Menten constants of the underpinning reactions, mostly $K_\delta$ and $K_D$. Remarkably, estimates of these two constants are in the range of $~0.1-\SI{1.0}{\micro \Molar}$, that is well within the range of \ca\ elevations expected for an astrocyte, whose average resting \ca\ concentration is reported to be $<\SI{0.15}{\micro \Molar}$ \citep{Zheng_Neuron2015}. This assures that activation of \plcd\ and \3k\ is effective only when intracellular \ca\ approaches to, or increases beyond $K_\delta$ and $K_D$, as expected by the occurrence of CICR.

\subsection{Signal integration}
The existence of different regimes of \ip3\ signaling shapes the time evolution of \ip3\ with respect to stimulation in a peculiar fashion. From \figref{fig:gchi-dynamics} (\textit{third panel}), it may indeed be noted that, starting from resting values, \ip3\ increases for each glutamate puff almost stepwise, till it reaches a peak (or threshold) concentration (normalized to~$\sim 1$) that triggers CICR, thereby triggering a \ca\ pulse-like elevation. This \ca\ elevation promotes \ip3\ degradation to some concentration between its peak and baseline values, in a sort of reset mechanism, leaving \ip3\ to increase back again to the CICR threshold until the next elevation. In between each \ca\ elevation, counting from the first one ending at $t\approx \SI{4}{s}$, we may appreciate how \ip3\ increases almost proportionally to the number of glutamate puffs, akin to an integrator of the stimulus. 

This may readily be proved by analytical arguments approximating, for simplicity, each glutamate puff occurring at $t_k$ by a Dirac's delta $\updelta(t-t_k)$, so that the external stimulus impinging on the astrocyte is modeled by $Y(t) = G\cdot\Delta\sum_k \updelta(t-t_k)$, where $G\cdot \Delta$ represents the glutamate concentration delivered in the time unit per puff (i.e. its dimensions are \si{\micro \Molar \cdot s}). Then, assuming that in between oscillations, intracellular \ca\ concentration is close to basal levels, i.e. $C\approx C_0$, with $C_0 < (\ll)\, K_{KC},\, K_\delta,\,K_3$ and $h\approx h_\infty$, it is possible to reduce equations~\ref{eq:GammaA} and~\ref{eq:gchi-I} to 
\begin{align}
\der{\Gamma_A}{t} &\approx -(O_N Y(t)+\Omega_N) \Gamma_A + O_N Y(t)\label{eq:GammaA-approx}\\
\der{I}{t}        &\approx -J_{5P} + J_\beta = - \Omega_{5P} I + O_\beta \Gamma_A \label{eq:I-approx}
\end{align}
Using the fact that for puffs delivered at rate $\nu$ the identity $\int_{t\,'}^{t\,''} \sum_k \updelta(t-t_k) \dd{t} = \nu (t\,'' - t\,')$ holds, we can solve \eqref{eq:GammaA-approx} for $\Gamma_A$ obtaining
\begin{align}
\Gamma_A(t) & = \int_{-\infty}^{t}O_N Y(t\,')\,e^{-\int_{t\,'}^{t}(\Omega_N+O_N Y(t\,''))\dd t\,''}\dd t\,' \nonumber\\
            & = \int_{-\infty}^{t}O_N Y(t\,')\,e^{-\Omega_N(t-t\,')}\,e^{-O_N\int_{t\,'}^{t}Y(t\,'')\dd t\,''}\dd t\,' \nonumber\\
            & = \int_{-\infty}^{t}O_N Y(t\,')\,e^{-\rpar{\Omega_N + O_N G \Delta \nu} (t-t\,')}\dd t\,' \nonumber\\
            & = O_N Y(t\,') \ast Z_{\Gamma_A}(t) \label{eq:GammaA-solution}
\end{align}
where ``$\ast$" denotes the convolution operator. It is thus apparent that the fraction of activated receptors $\Gamma_A(t)$ is an integral transform of the stimulus $Y(t)$ by convolution with the kernel~$Z_{\Gamma_A}(t)$. Specifically, $Z_{\Gamma_A}(t)$ may be regarded as the fraction of astrocyte receptors stimulated by one extracellular glutamate puff -- or equivalently, by synaptic release triggered by an action potential --, and characterizes the encoding of the stimulus by the astrocyte via its activated receptors.

The \ip3\ signal resulting from the activated receptors then evolves according to
\begin{align}
I(t) & = \int_{-\infty}^{t} O_\beta \Gamma_A (t\,')\,e^{-\int_{t\,'}^{t}\Omega_{5P}\dd t\,''}\dd t\,' = \int_{-\infty}^{t} O_\beta \Gamma_A (t\,')\,e^{-\Omega_{5P}(t-t\,')}\dd t\,'\nonumber\\
     & = O_\beta\Gamma_A (t) \ast Z_{I}(t) \label{eq:I-solution}
\end{align}
That is the \ip3\ signal is also an integral transform of the input stimuli through the fraction of activated receptors $\Gamma_A(t)$, by convolution with the kernel~$Z_{I}(t)=e^{-\Omega_{5P}t}$. In particular, experimental evidence hints that the rate constant $\Omega_{5P}$ is often small compared to the rate of incoming stimulation (\appref{app:Parameters}), so that $Z_{I}(t)\approx 1$. In this case then, \eqref{eq:I-solution} predicts that $I(t)\approx \int_{-\infty}^{t}O_\beta \Gamma_A(t\,')\dd t\,'$, namely that the \ip3\ signal effectively corresponds to the integral of the fraction of activated astrocyte receptors.\\[1.5ex]

\begin{figure}[!tp]
	\begin{subfigure}[t]{0.49\textwidth}
		\caption{}
		\includegraphics[width=1.0\textwidth]{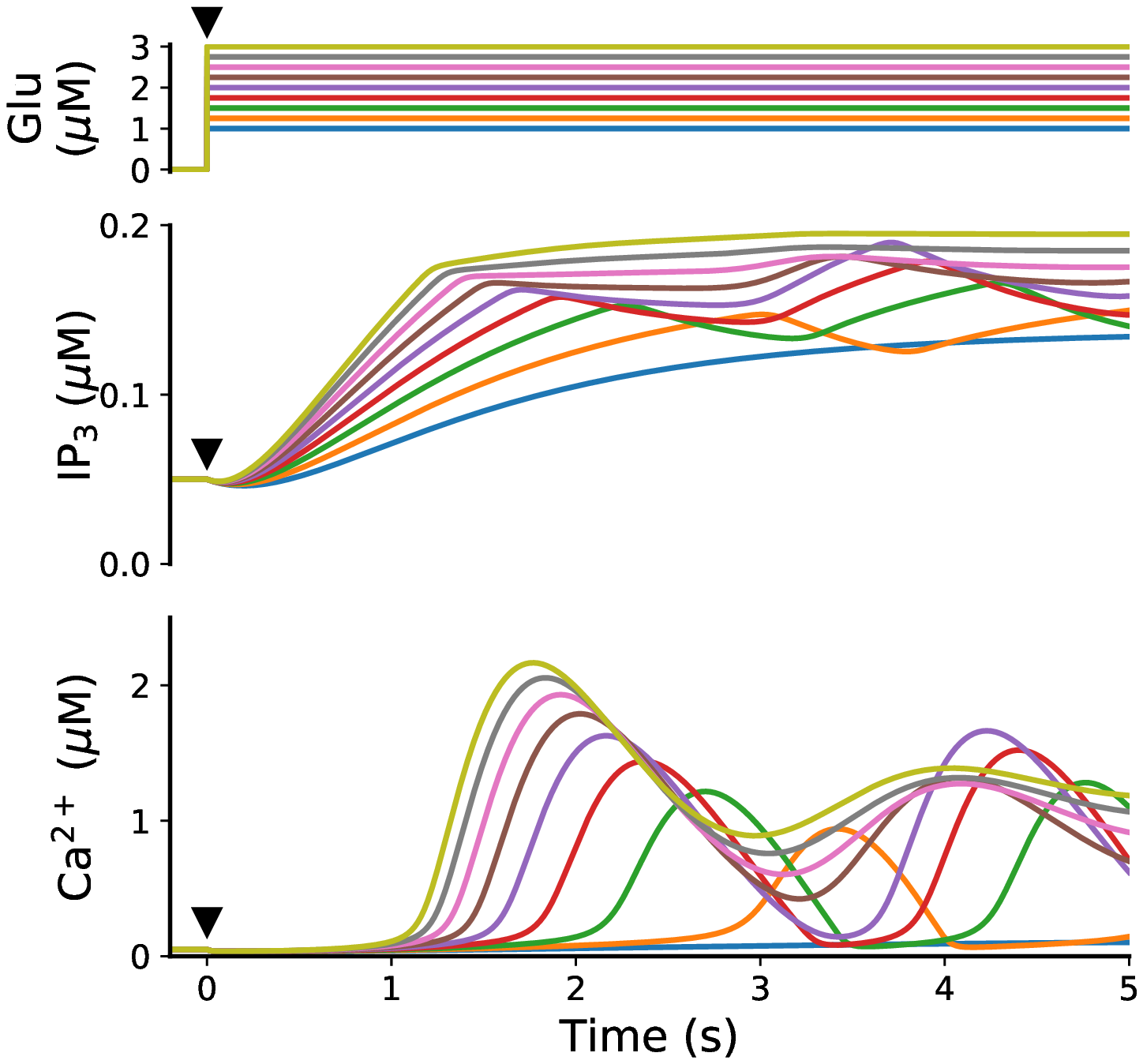}		
	\end{subfigure}
	\begin{subfigure}[t]{0.49\textwidth}
		\caption{}
		\includegraphics[width=0.83\textwidth]{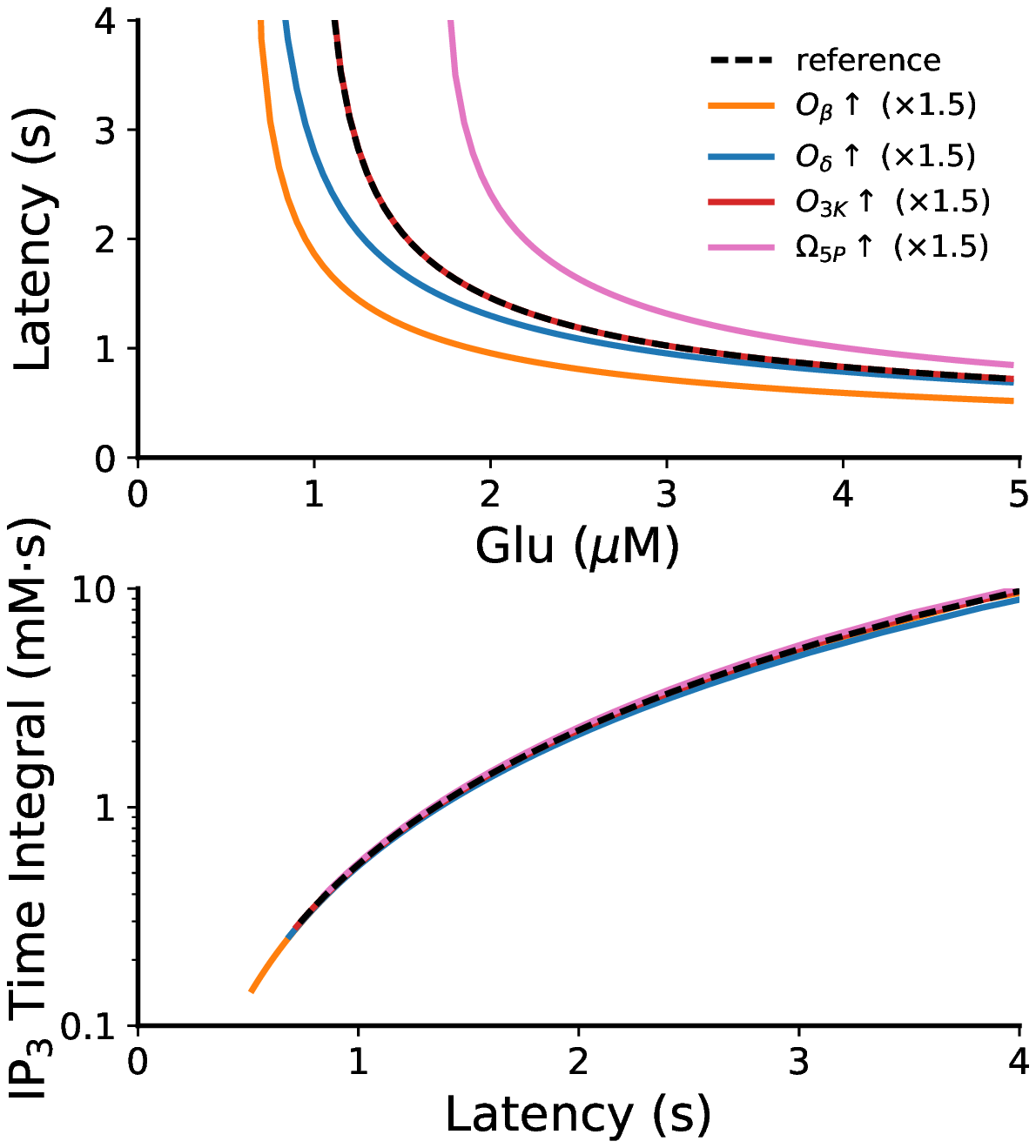}		
	\end{subfigure}
	\caption{Threshold for CICR. \textbf{A}~(\textit{top panel}) Step increases of extracellular glutamate (\textit{color coded}) and resulting~\ip3\ (\textit{central panel}) and \ca\~dynamics (\textit{bottom panel}) in a~\gchi\ astrocyte model. \textit{Black marks} at $t=0$ denote stimulus onset. \textbf{B}~(\textit{top panel}) Latency for the onset of CICR as a function of the applied glutamate concentration for the \ca\ traces in~\textbf{A} (\textit{black dashed curve}), as well as for 50\% increases in the rate of \plcb\ ($O_\beta$), \plcd\ ($O_\delta$), \3k\ ($O_{3K}$) and \5p\ ($\Omega_{5P}$) respectively. Emergence of CICR was detected for $\der{C}{t} \ge \SI{0.5}{\micro \Molar / s}$. (\textit{bottom panel})~Integral of \ip3\ concentration as a function of the latency values computed in the top panel. This integral is a better estimator of CICR threshold than the sole \ip3\ concentration. Model parameters as \figref{fig:gchi-dynamics}.}
	\label{fig:gchi-threshold}
\end{figure}

It is also worth understanding the nature of the threshold concentration that \ip3\ must reach in order to trigger CICR. In the \gchi\ model, based on the Li-Rinzel description of CICR, this threshold may be not well-defined and generally varies with the parameter choice as well as with the shape and amplitude of the delivered stimulation \citep{DePitta_JOBP2009}. Consider for example \figref{fig:gchi-threshold}A where the \ca\ response of an astrocyte (\textit{bottom panel}) is simulated for different \textit{color-coded} step increases of extracellular glutamate (\textit{top panel}). It may be noted that CICR, reflected by one or multiple \ca\ pulse-like increases, is triggered by glutamate concentrations greater or equal to the \textit{orange trace}. However, the \ip3\ threshold for CICR (\textit{central panel}) appears to grow with the extracellular glutamate concentration. This is reflected by the first 'knee' of the \ip3\ curves which reaches progressively higher values of \ip3\ concentration as extracellular glutamate increases from \textit{orange} to \textit{lime} levels. At the same time, as shown by the \textit{black dashed curve} in the \textit{top panel} of \figref{fig:gchi-threshold}B, the latency for emergence of CICR since stimulus onset (\textit{black marks} at $t=0$) decreases. This can be explained by equations~\ref{eq:GammaA-approx} and~\ref{eq:I-approx}, noting that, while larger glutamate concentrations promote larger receptor-mediated \ip3\ production, this increased production is also counteracted by faster degradation by \5p, since this latter linearly increases with \ip3. Thus while larger \ip3\ production assures shorter delays in the onset of CICR, a larger \ip3\ level must be reached to compensate for its faster degradation.

The \textit{top panel} of \figref{fig:gchi-threshold}B further illustrates how the latency period for CICR onset depends on the activity of the different enzymes regulating \ip3\ production and degradation. Here the different \textit{colored curves} were obtained repeating the simulations of \figref{fig:gchi-threshold}A for a 50\% increase of the activity respectively of \plcb\ (\textit{orange trace}), \plcd\ (\textit{blue trace}), \3k\ (\textit{red trace}) and \5p\ (\textit{violet trace}). In agreement with our previous analysis, \plcb\ and \5p\ have the largest impact on respectively reducing or increasing the latency period, given that they are the main enzymes at play in \ip3\ signaling before CICR onset. The effect of an increase of \ip3\ production by \plcd\ is instead mainly significant for low glutamate concentrations, such that they could promote an activation of this enzyme that is comparable to that of \plcb. Conversely, \3k\ does not have any role in the control of CICR latency since its activation effectively requires CICR to onset first.

The variability of \ip3\ concentrations attained to trigger CICR by different glutamate concentrations, and its correlation with the latency for CICR onset, suggest that the mere \ip3\ concentration is not an effective indicator of the CICR threshold, rather we should consider instead the total \ip3\ amount produced in the astrocyte cytosol during the latency period that precedes CICR onset, that is the integral in time of \ip3\ concentration during such period. This is exemplified in the \textit{bottom panel} of \figref{fig:gchi-threshold}B where such integral is plotted as a function of the different latency values computed in the \textit{top panel}. It may be appreciated how this integral is essentially similar for different enzyme expressions (\textit{colored curves}) yet associated with the same latency value. 

Taken together these results put emphasis on the crucial role exerted by \ip3\ signaling in the genesis of agonist-mediated \ca\ elevations. In particular they suggest that the expression of different enzymes responsible of \ip3\ production and degradation, which is likely heterogeneous across an astrocyte, could locally set different requirements for integration and encoding of external stimuli by the same cell.

\subsection{Role of cPKCs and beyond}
Different mechanisms of production and degradation of \ip3\ are only one example of the possible many signaling pathways that could shape the nature of \ca\ signaling in astrocytes. There is also compelling evidence in vitro that shape and duration of \ca\ oscillations could be controlled by astrocyte receptor phosphorylation by \ce{cPKCs} \citep{CodazziTeruelMeyer2001}. To better understand this aspect of astrocyte \ca\ signaling, we relax the quasi steady-state approximation on \ce{cPKC} phsophorylation and thus rewrite \eqref{eq:R*-ode-qssa} as
\begin{equation}
\der{\Gamma_A}{t} = O_N \ce{[A]$^{n}$}\,\rpar{1-\Gamma_A} - \rpar{\Omega_N + O_K P}\Gamma_A \label{eq:GammaA-cpkc}
\end{equation}
where $P$ denotes the \cpkc\ concentration at the receptors' site. This in turn, requires to also consider a description of \cpkc\ dynamics, whereby at least two additional equations in the \gchi\ model must be included: one that takes into account $P$ dynamics, but also a further one that describes \dag\ dynamics ($D$), which is responsible for \ce{cPKC} activation by \ca-dependent translocation of the inactive kinase to the plasma membrane \citep{Oancea_Cell1998}.

By~QSSA, the quantity of \cpkc\ is conserved during receptor phosphorylation in reaction~\ref{rc:R-PKC*-reaction}. In this fashion, \cpkc\ production and degradation are only controlled by the pair of reactions~\ref{rc:PKC*-reaction-ca} and~\ref{rc:PKC*-reaction-dag}. On the other hand, taking into account from \secref{sec:PLCb-production} that production of \cpkc\ depends on the availability of the \ca-bound kinase complex \ce{cPKC$\,'$}, we may assume at first approximation that reaction~\ref{rc:PKC*-reaction-ca} for \ca-binding to the kinase is at equilibrium, i.e. \ce{[cPKC$\,'$] = [cPKC]_T \Hill{1}{C}{K_{KC}}}. Accordingly, we can consider \cpkc\ dynamics to be driven simply by reaction~\ref{rc:PKC*-reaction-dag}, i.e.
\begin{align}
\der{P}{t} &= J_{KP} - J_{KD}\nonumber\\
           &= O_{KD}\ce{[cPKC$\,'$]}\cdot D - \Omega_{KD} P \nonumber\\
           &= O_{KD}\ce{[cPKC]_T}\Hill{1}{C}{K_{KC}}\cdot D - \Omega_{KD} P \nonumber\\
           &\equiv O_{KD}\Hill{1}{C}{K_{KC}}\cdot D - \Omega_{KD} P \label{eq:cPKC}
\end{align}
where we re-defined $O_{KD}\leftarrow O_{KD}\ce{[cPKC]_T}$ as the maximal rate of \cpkc\ production (in \si{\micro \Molar s^{-1}}).

To model \dag\ dynamics we start instead from the consideration that PLC isoenzymes hydrolyze \pip2\ into one molecule of \ip3 and one of \dag, so that \dag\ production coincides with that of \ip3\ \citep[][and see also \figref{fig:ip3-degradation}B]{BerridgeIrvine_Nature1989}. Yet, only part of this produced \dag\ is used to activate \ce{cPKC}, while the rest is mainly degraded by diacylglycerol kinases (DAGKs) into phosphatidic acid \citep{Carrasco_TiBS2007} and, to a minor extent, by diacylglycerol lipases (DAGLs) into 2-arachidonoylglycerol (2-AG), although this latter pathway has only been linked to some types of metabotropic receptors in astrocytes \citep{BrunerMurphy_JNC1990,Giaume_PNAS1991,Walter_JN2004}. Other pathways of use of \dag\ are also possible in principle, inasmuch as \dag\ is a key molecule in the cell's lipid metabolism and a basic component of membranes. Nonetheless there is evidence that \dag\ levels are strictly regulated within different subcellular compartments, and \dag\ generated by GPCR stimulation is not usually consumed for metabolic purposes \citep{van-der-Bend1994,Carrasco_TiBS2007}.

DAGK activation reflects the sequence of \ca\-mediated translocation, \dag\ binding and activation that is also required for cPKCs, so the two reactions may be thought to be characterized by similar kinetics, yet with an important difference. Sequence analysis of DAGK$\upalpha,\,\upgamma$ -- the two isoforms of DAGKs most likely involved in astrocytic GPCR signaling \citep{Dominguez_CD2013} -- reveals in fact the existence of two EF-hand motifs characteristics of \ca-binding and two C1 domains for \dag\ binding \citep{Merida_BJ2008}. In this fashion, a Hill exponent of~2 instead of~1 as in \eqref{eq:cPKC} must be considered for the DAGK activating reaction, so that DAGK-mediated DAG degradation can be modeled by
\begin{equation}
J_D = O_{D} \Hill{2}{C}{K_{DC}}\Hill{2}{D}{K_{DD}}
\end{equation}
Finally, to take into account other mechanisms of \dag\ degradation ($J_A$), including but not limited to DAGLs, we assume a linear degradation rate, i.e. $J_A = \Omega_D D$. This is a crude approximation insofar as DAGL, could also be activated in a \ca-dependent fashion \citep{Rosenberger_Lipids2007}. Nonetheless, the complexity of the molecular reactions likely involved in these other pathways of DAG degradation would require to consider additional equations in our model which are beyond the scope of this chapter. The reader who is interested in these further aspects, may refer to \citet{Cui_eLife2016} for a possible modeling approach. For the purposes of our analysis instead, we will consider the following equation for \dag\ dynamics:
\begin{align}
\der{D}{t} &= J_\beta + J_\delta - J_{KP} - J_D - J_A \nonumber\\
           &= O_\beta \Gamma_A + O_{\delta}\Hill{2}{C}{K_\delta}\rpar{1-\Hill{1}{I}{\kappa_\delta}} + \nonumber\\
           &\phantom{=} -O_{KD} \Hill{1}{C}{K_{KC}}\cdot D - O_{D} \Hill{2}{C}{K_{DC}}\Hill{2}{D}{K_{DD}} - \Omega_D D \label{eq:DAG}
\end{align}

\begin{figure}[!tp]
	\begin{subfigure}[t]{0.285\textwidth}
		\caption{}
		\includegraphics[width=1.0\textwidth]{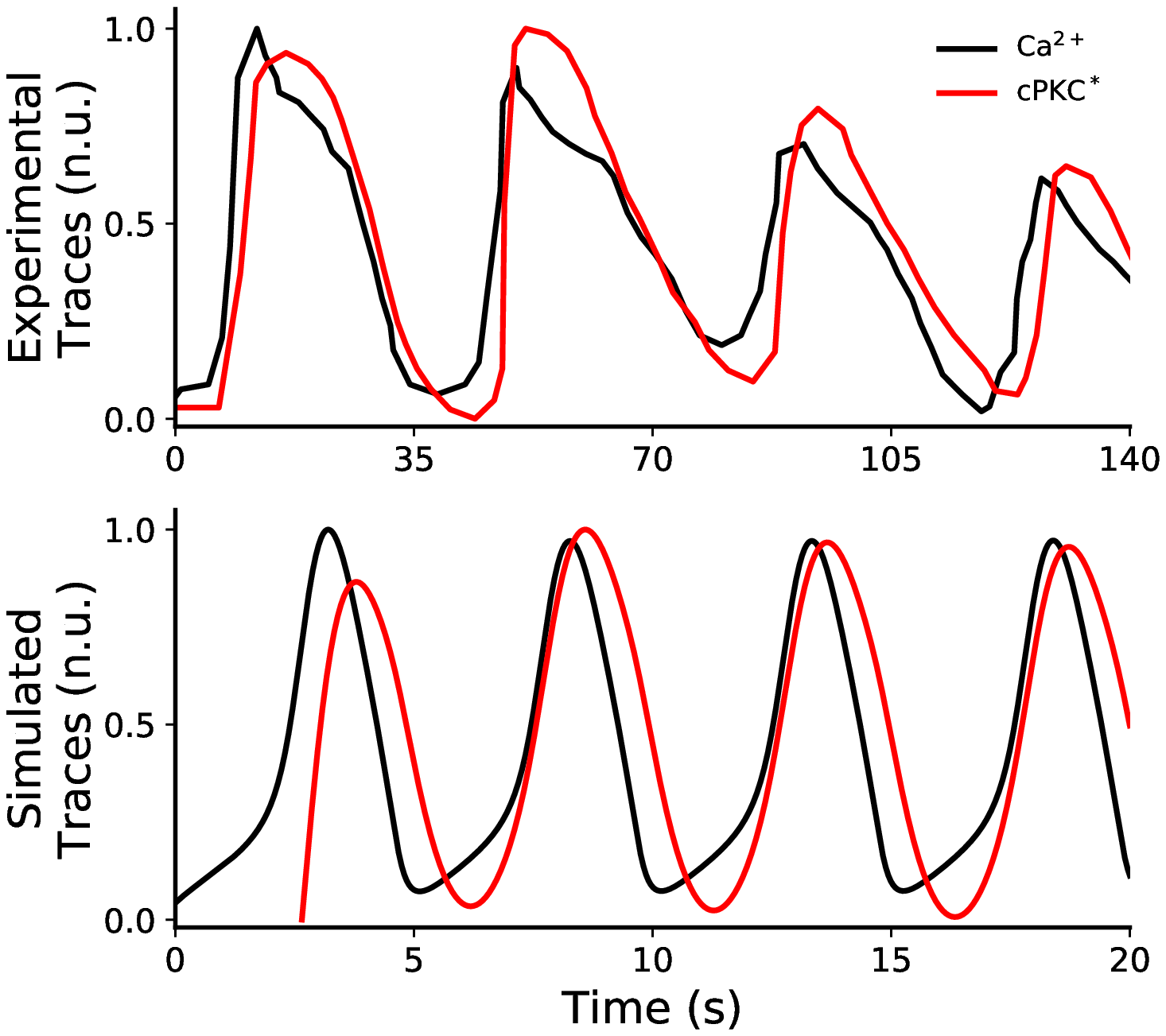}		
	\end{subfigure}
	\begin{subfigure}[t]{0.35\textwidth}
		\caption{}
		\includegraphics[width=1.0\textwidth]{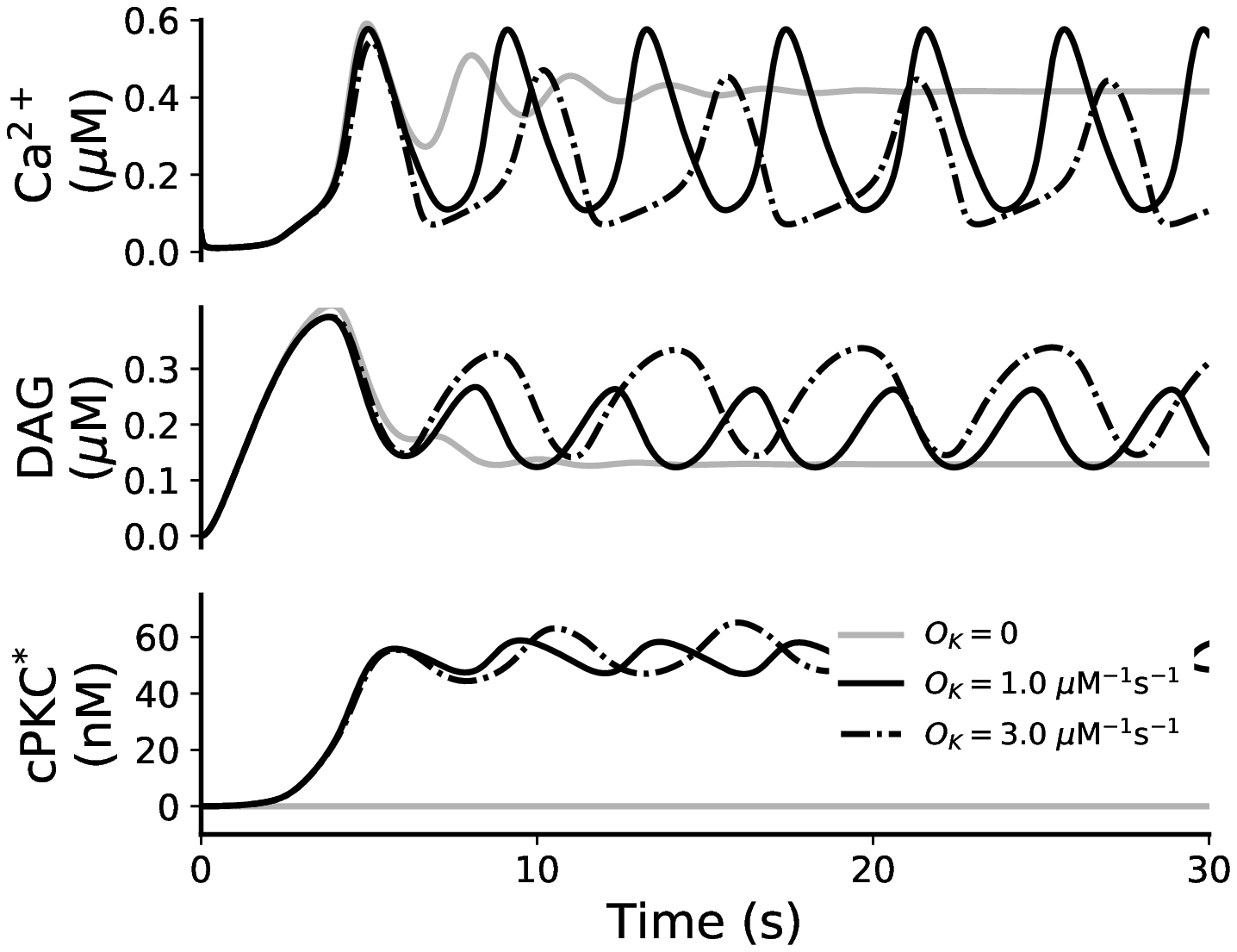}		
	\end{subfigure}
	\begin{subfigure}[t]{0.35\textwidth}
		\caption{}
		\includegraphics[width=1.0\textwidth]{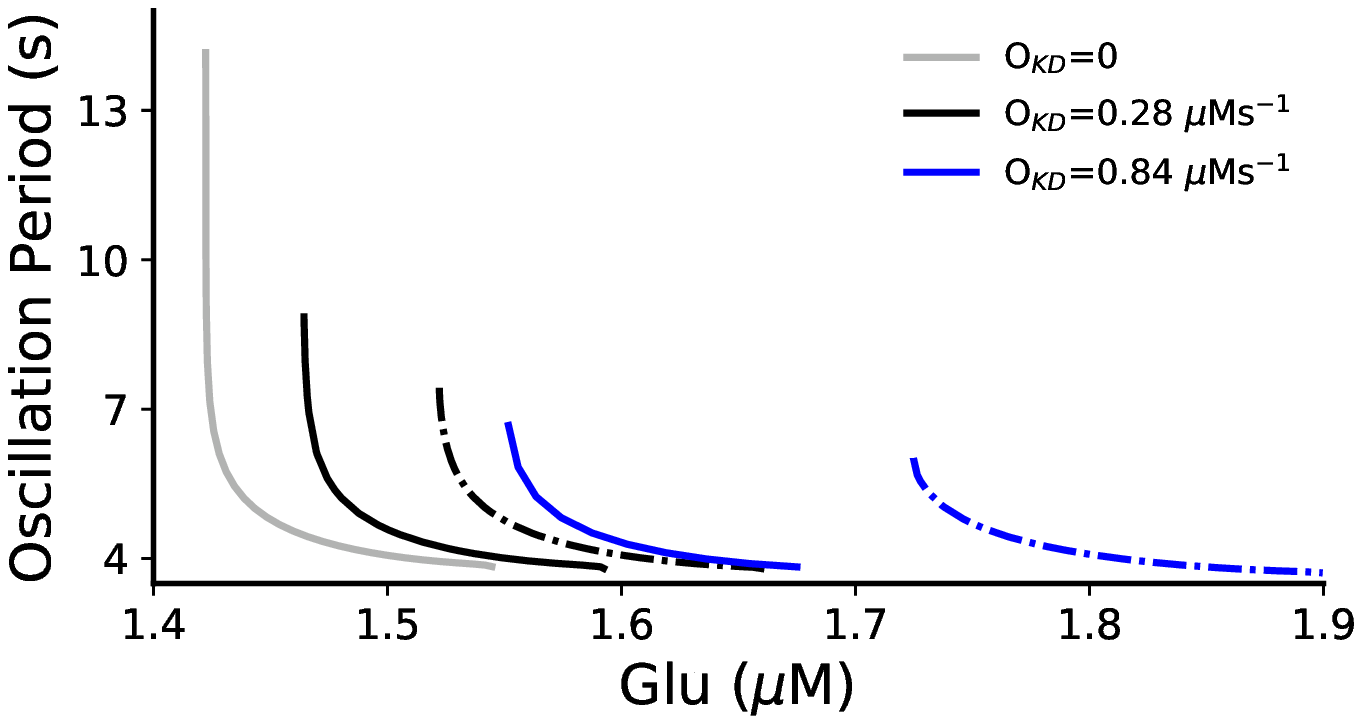}		
	\end{subfigure}
	\caption{Regulation of \ca\ oscillations by \ce{cPKC}. \textbf{A}~(\textit{top panel}) Comparison between experimental traces for \ca\ (\textit{black}) and \cpkc\ (\textit{red}) originally recorded in cultured astrocytes by \citet{CodazziTeruelMeyer2001} and simulations (\textit{bottom panel}). Despite quantitive differences in the shape and period of oscillations, the model can reproduce the essential correlation and phase shift between \ca\ and \cpkc\ dynamics observed in experiments. \ca\ and \cpkc\ oscillations were triggered assuming an extracellular glutamate concentration of~\SI{1.48}{\micro \Molar}, and were normalized according to their maximum excursion: $C_0=\SI{0.04}{\micro \Molar}$, $P_0=\SI{48}{\nano \Molar}$, $\hat{C}=\SI{0.49}{\micro \Molar}$ and $\hat{P}=\SI{65}{\nano \Molar}$. \textbf{B}~DAG and \cpkc\ dynamics associated with two different rates of receptor phosphorylation by \ce{cPKC} ($O_K$, \textit{black traces}) in response to a step increase of extracellular glutamate (\SI{1.55}{\micro \Molar} at $t=0$). In the absence of  receptor phosphorylation (\textit{gray traces}), \ca\ oscillations would vanish due to saturating intracellular \ip3\ levels ensued from large receptor activation. \textbf{C}~Period of \ca\ oscillations as a function of extracellular glutamate concentration. Receptor phosphorylation by \ce{cPKC} critically controls the oscillatory range (\textit{black} and \textit{blue curves}) with respect to the scenario without \ce{cPKC} activation (\textit{gray curve}). Higher glutamate concentrations are required to trigger oscillations for larger rates of \dag-dependent \ce{cPKC} activation~($O_{KD}$). Parameters as in \tabref{tab:Model-Parameters} except for $\Omega_C=\SI{6.207}{s^{-1}},\, \Omega_L=\SI{0.01}{s^{-1}},\, O_\beta=\SI{1}{\micro \Molar s^{-1}}$.}
	\label{fig:gchidp}
\end{figure}

\figref{fig:gchidp}A shows a comparison of experimental \ca\ and \cpkc\ traces with those reproduced by the \gchi\ model including equations~\ref{eq:cPKC} and~\ref{eq:DAG}. For inherent limitations of the Li-Rinzel description of the gating kinetics of \ip3Rs, which fails to describe these receptors' open probability for large \ca\ concentration (\figref{fig:chi}) and predicts fast rates of receptor de-inactivation ($O_2 / d_2$, \tabref{tab:Model-Parameters}), the \gchi\ model cannot generate \ca\ peaks as large as those experimentally observed and shown here. Nonetheless we would like to emphasize how our model qualitatively matches experimental \ca-dependent \cpkc\ dynamics, accurately reproducing the phase shift between \ca\ and \cpkc\ oscillations. This phase shift is critically controlled by the constant $K_{KC}$ for \ca\ binding to the kinase, along with the rates of \cpkc\ production vs. degradation, i.e. $O_{KD}$ vs. $\Omega_{KD}$ (\eqref{eq:cPKC}), and the rate of receptor phosphorylation $O_K$ (\eqref{eq:GammaA-cpkc}). 

\figref{fig:gchidp}B further reveals the role of these rate constants in the control of \ca\ oscillations. In this figure, we simulated the astrocyte response for a step increase of $\sim \SI{1.5}{\micro \Molar}$ extracellular glutamate, starting from resting conditions, both in the absence of kinase-mediated receptor phosphorylation (\textit{gray trace}) and in the presence of it, for two different $O_{K}$ rate values (\textit{black traces}). It may be noted how receptor phosphorylation by \ce{cPKC} can rescue \ca\ oscillations that otherwise would vanish by saturating intracellular \ip3\ concentrations ensuing from large receptor activation. This activation indeed is decreased by \cpkc\ according to \eqref{eq:GammaA-cpkc}, thereby regulating intracellular \ip3\ within the range of \ca\ oscillations. Nonetheless, as the rate of receptor phosphorylation increases (\textit{dash-dotted trace}), the period of oscillations appears to slow down and oscillations even fail to emerge, if the supply of \cpkc\ results in a phosphorylation rate of astrocyte receptors that exceeds their agonist-mediated activation (results not shown).

These considerations can be explained considering the period of \ca\ oscillations as a function of the extracellular glutamate concentration. As shown in \figref{fig:gchidp}C, \ce{cPKC}-mediated receptor phosphorylation shifts (\textit{black curves}) the range of glutamate concentrations that trigger \ca\ oscillations to higher values than those otherwise expected in the absence of it (\textit{gray curve}). In particular, and in agreement with experimental findings \citep{CodazziTeruelMeyer2001}, the exact value of the rate $O_K$ for receptor phopshorylation sets the entity of this shift, accounting either for \ca\ oscillations of period longer than without receptor phosphorylation, or for the requirement of larger glutamate concentrations to observe such oscillations. This is respectively reflected by the portions of the \textit{black curves} that are within the range of extracellular glutamate concentrations of the \textit{gray curve}), and those that instead are not. On the other hand, longer-period oscillations in the presence of receptor phosphorylation are likely to be observed as long as the rate of \cpkc\ activation by DAG ($O_{KD}$) is below some critical value. A three-fold increase of this rate indeed requires glutamate concentrations beyond those needed in the absence of receptor phosphorylation to trigger oscillations, regardless of the $O_K$ value at play (\textit{blue curves}). In this scenario in fact, the large supply of \cpkc, resulting from the high $O_{KD}$ value, favors phosphorylation of receptors while hindering intracellular buildup of \ip3\ to trigger CICR. This in turn requires a larger recruitment of astrocyte receptors by larger agonist concentrations to evoke \ca\ oscillations.

\section{Conclusions}
The modeling arguments introduced in this chapter overall suggest a great richness in the possible modes whereby astrocytes could translate extracellular stimuli into intracellular \ca\ dynamics. These modes are brought forth by a complex network of biochemical reactions that is exquisitely nonlinearly coupled with \ca\ dynamics through different second messengers, among which \ip3\ and possibly DAG could play a paramount signaling role. In particular, the regulation of different regimes of \ip3\ production and degradation by \ca\ in parallel with the differential regulation by this latter and DAG of the activities of cPKCs and DAGKs opens to the scenario of the existence of different regimes of signal transduction that a single astrocyte could multiplex towards different intracellular targets depending on different local conditions of neuronal activity. 

An interesting implication emerging from our analysis of the regulation of the period of \ca\ oscillations by cPKCs and DAG-related lipid signals is the possibility that these pathways, which could be crucially linked with inflammatory responses underpinning reactive astrocytosis \citep{Brambilla_BJP1999,Griner_NRC2007}, could be found at different operational states, akin to what suggested for proinflammatory cytokines like~TNF$\upalpha$ \citep{Santello_TiNS2012}. In our analysis for example, intermediate activation of~cPKC activity could promote \ca\ oscillations at physiological rates, while an increase of it could exacerbate fast, potentially inflammatory \ca\ responses \citep{Sofroniew_AN2010}. 

Similar arguments also hold for \ip3\ signaling. Calcium-dependent \ip3\ production by \plcd\ and \plcb\ (via cPKC) could modulate the rate of integration of synaptic stimuli and thus dictate the threshold synaptic activity triggering CICR. On the other hand, the existence of different regimes of \ip3\ degradation could be responsible for different cutoff frequencies of synaptic release, beyond which integration of external stimuli by the cells could cease. In particular, this cutoff frequency could be mainly set by \5p\ during low synaptic activity, possibly associated with low intracellular \ca\ levels, while be dependent on \3k\ in regimes of strong astrocyte \ca\ activation, and thus ultimately depend on the history of activation of the astrocyte. The following chapter looks closely at some of these aspects, focusing in particular, on the role of different \ip3\ degradation regimes in the genesis and shaping of \ca\ oscillations.

\newpage
\begin{appendices}
\renewcommand\thetable{\thesection\arabic{table}}
\section{Arguments of chemical kinetics}\label{app:chem-kinetics}
\subsection{The Hill equation}\label{app:Hill-function}
In biochemistry, the binding reaction of~$n$ molecules of a ligand~$L$ to a receptor macromolecule~$R$, i.e.,
\begin{equation}\label{eq:binding-reaction}
\ce{R + $n$L <=>T[$k_{f}$][$k_{b}$] RL_n}
\end{equation}
can be mathematically described by the differential equation
\begin{equation}\label{eq:binding-ode}
\der{\ce{[RL_n]}}{t}=k_{f}\ce{[R][L]^n}-k_{b}\ce{[RL_n]}
\end{equation}
where~$k_{f}$,~$k_{b}$ denote the forward (binding) and backward (unbinding) reaction rates respectively. At equilibrium, 
\begin{equation}\label{eq:binding-equilibrium}
0 = k_{f}\ce{[R][L]^n}-k_{b}\ce{[RL_n]} \Rightarrow \ce{[RL_n]}=\frac{\ce{[R][L]^n}}{K_d}
\end{equation}
where~$K_{d}=k_{b}/k_{f}$ is the \emph{dissociation constant} of the binding reaction~\ref{eq:binding-reaction}. Then, the fraction of bound receptor macromolecules with respect to the total receptor macromolecules can be expressed by the Hill equation \citep{StryerBiochemistryBOOK}
\begin{equation}\label{eq:Hill-equation}
\frac{\textrm{Bound}}{\textrm{Total}} = \frac{\ce{[RL_{n}]}}{\ce{[R] + [RL_{n}]}} = \dfrac{\dfrac{\ce{[L]^n}}{K_{d}}}{\dfrac{[L]^{n}}{K_{d}}+1} = \dfrac{\ce{[L]^n}}{\ce{[L]^n}+K_{d}} = \dfrac{[L]^{n}}{[L]^{n}+K_{0.5}^{n}}=\Hill{n}{\ce{[L]}}{K_{0.5}}
\end{equation}
where the function~\Hill{n}{\ce{[L]}}{K_{0.5}} denotes the sigmoid (Hill) function~$\ce{[L]^n / ([L]^n + K_{0.5}^n})$, and $K_{0.5} = \sqrt[n]{K_{d}}$ is the receptor \emph{affinity} for the ligand $L$, and corresponds to the ligand concentration for which half of the receptor macromolecules are bound (i.e.~the midpoint of the \Hill{n}{\ce{[L]}}{K_{0.5}} curve). The sigmoid shape of~\Hill{n}{\ce{[L]}}{K_{0.5}} denotes \emph{saturation kinetics} in the binding reaction~\ref{eq:binding-reaction}, that is, for~$\ce{[L]}\gg K_{0.5}$ almost all the receptor molecules are bound to the ligand, so that the fraction of bound receptor molecules does not essentially change for an increase of~$[L]$.

The coefficient~$n$, also known as \emph{Hill coefficient}, quantifies the cooperativity among multiple ligand binding sites. A Hill coefficient~$n>1$ denotes \emph{positively cooperative binding}, whereby once one ligand molecule is bound to the receptor macromolecule, the affinity of the latter for other ligand molecules increases. Conversely, a value of~$n<1$ denotes \emph{negatively cooperative binding}, namely when binding of one ligand molecule to the receptor decreases the affinity of the latter to bind further ligand molecules. Finally, a coefficient~$n=1$ denotes completely \emph{independent binding} when the affinity of the receptor to ligand molecules is not affected by its state of occupation by the latter.

For unimolecular reactions,~$n=1$ coincides with the number of binding sites of the receptor. For multimolecular reactions involving~$\eta > 1$ ligand molecules instead, the Hill coefficient in general, only loosely estimates the number of binding sites, being~$n\le \eta$ \citep{Weiss1997}. This follows from the hypothesis of total allostery that is implicit in the reaction~\ref{eq:binding-reaction}, whereby the Hill function is a very simplistic way to model cooperativity. It describes in fact the limit case where affinity is~0 if no ligand is bound, and~infinite as soon as one receptor binds. That is, only two states are possible: free receptor and receptor with all ligand bound. More realistic descriptions are available in literature, such as for example the Monod--Wyman--Changeux~(MWC) model, but they yield much more complex equations and more parameters \citep{Changeux_Science2005}.

\subsection{The Michaelis-Menten model of enzyme kinetics}\label{app:Michaelis-Menten-kinetics}
The Michaelis-Menten model of enzyme kinetics is one of the simplest and best-known models to describe the kinetics of enzyme-catalyzed chemical reactions. In general enzyme-catalyzed reactions involve an initial binding reaction of an enzyme~\ce{E} to a substrate~\ce{S} to form a complex~\ce{ES}. The latter is then converted into a product~\ce{P} and the free enzyme by a further reaction that is mediated by the enzyme itself and can be quite complex and involve several intermediate reactions. However, there is typically one rate-determining enzymatic step that allows this reaction to be modeled as a single catalytic step with an apparent rate constant~$k_{\mathrm{cat}}$. The resulting kinetic scheme thus reads
\begin{equation}\label{eq:Michaelis-Menten-reaction}
\ce{E + S <=>T[$k_{f}$][$k_{b}$] ES ->T[$k_{\mathrm{cat}}$] P + E}
\end{equation}
By law of mass action, the above kinetic scheme gives rise to~4 differential equations \citep{StryerBiochemistryBOOK}:
\begin{subequations}\label{eq:Michaelis-Menten-equations}
\begin{align}
\der{\ce{[S]}}{t} & = -k_{f}\ce{[E][S]} + k_{b}\ce{[ES]}\\
\der{\ce{[E]}}{t} & = -k_{f}\ce{[E][S]} + k_{b}\ce{[ES]} + k_{\mathrm{cat}}\ce{[ES]} \label{eq:Michaelis-Menten-E}\\
\der{\ce{[ES]}}{t}& = k_{f}\ce{[E][S]} - k_{b}\ce{[ES]} - k_{\mathrm{cat}}\ce{[ES]} \label{eq:Michaelis-Menten-ES}\\
\der{\ce{[P]}}{t} & = k_{\mathrm{cat}}\ce{[ES]} \label{eq:Michaelis-Menten-P}
\end{align}
\end{subequations}
In the Michaelis-Menten model the enzyme is a catalyst, namely it only facilitates the reaction whereby~\ce{S} is transformed into~\ce{P}, hence its total concentration~\ce{[E]_T=[E] + [ES]} must be preserved. This is indeed apparent by the sum of the second and the third equations above, since:~$\der{(\ce{[E] + [ES]})}{t}=\der{\ce{[E]_T}}{t}=0\Rightarrow \ce{[E]_T} = \textrm{const}$.

The system of equations~\ref{eq:Michaelis-Menten-equations} can be solved for the products~\ce{P} as a function of the concentration of the substrate~\ce{[S]}. A first solution assumes instantaneous chemical equilibrium between the substrate~\ce{S} and the complex~\ce{ES}, i.e.~$\der{\ce{[S]}}{t}=0$, whereby the initial binding reaction can be equivalently described by a Hill equation \citep{KeenerSneyd_2008_Book}, i.e.,
\begin{equation}\label{eq:Michaelis-Menten-ChemEq}
\frac{\ce{[ES]}}{\ce{[E]_T}} = \frac{\ce{[S]}}{\ce{[S]} + K_{d}} \Rightarrow \ce{[ES]} = \frac{\ce{[E]_T [S]}}{\ce{[S]} + K_{d}}
\end{equation}
Alternatively, the \emph{quasi-steady-state assumption}~(QSSA) that~\ce{[ES]} does not change on the time scale of product formation can be made, so that~$\frac{d}{dt}\ce{[ES]} = 0 \Rightarrow k_{f}\ce{[E][S]} = k_{b}\ce{[ES]} + k_{\mathrm{cat}}\ce{[ES]}$ \citep{KeenerSneyd_2008_Book}, and
\begin{align}\label{eq:Michaelis-Menten-QSSA}
k_{f}\ce{[E][S]} = k_{b}\ce{[ES]} + k_{\mathrm{cat}}\ce{[ES]} & \Rightarrow k_{f}
\ce{\rpar{\ce{[E]_T} - \ce{[ES]}}[S]} = k_{b}\ce{[ES]} + k_{\mathrm{cat}}\ce{[ES]} \nonumber\\
& \Rightarrow k_{f}\ce{[E]_T [S]} = \rpar{k_{f}\ce{[ES]}\ce{[S]}+k_{b}\ce{[ES]} + k_{\mathrm{cat}}\ce{[ES]}} \nonumber\\
& \Rightarrow \ce{[ES]} = \ce{[E]_T}\frac{\ce{[S]}}{\ce{[S]}+K_{\mathrm{M}}}
\end{align}
where~$K_{\mathrm{M}}=\rpar{k_{b}+k_{\mathrm{cat}}}/k_{f}$ is the \emph{Michaelis-Menten constant} of the reaction which quantifies the affinity of the enzyme to bind to the substrate.

Regardless of the hypothesis made to find an expression for~\ce{[ES]}, the rate~$v_{P}$ of production of~\ce{P} can be always written as
\begin{equation}\label{eq:Michaelis-Menten-Hill}
v_{P} = \der{\ce{[P]}}{t} = k_{\mathrm{cat}}\ce{[ES]} = k_{\mathrm{cat}}\ce{[E]_T}\frac{\ce{[S]}}{\ce{[S]}+K_{0.5}} = v_{max}\frac{\ce{[S]}}{\ce{[S]}+K_{0.5}}
\end{equation}
where~$v_{max}=k_{\mathrm{cat}}\ce{[E]_T}$ is the maximal rate of production of~\ce{P} in the presence of enzyme saturation, when all the available enzyme takes part in the reaction; and the affinity constant~$K_{0.5}$ equals the dissociation constant~$K_{d}$ of the initial binding reaction in the chemical equilibrium approximation~(\eqref{eq:Michaelis-Menten-ChemEq}), or the Michaelis-Menten constant in the~QSSA~(\eqref{eq:Michaelis-Menten-QSSA}).

An important corollary of the Michaelis-Menten model of enzyme kinetics is that the fraction of the total enzyme that forms the intermediate complex~\ce{ES} can be expressed by a Hill equation of the type
\begin{equation}\label{eq:Michaelis-Menten-Hill}
\frac{\ce{[ES]}}{\ce{[E]_T}} = \frac{\ce{[S]}}{\ce{[S]}+K_{0.5}} = \Hill{1}{[S]}{K_{0.5}}
\end{equation}
and~$K_{0.5}$ can be regarded as the half-saturating substrate concentration of the reaction. Similarly, the effective reaction rate~$v_{P}$ (\eqref{eq:Michaelis-Menten-Hill}) is proportional to the maximal reaction rate by a Hill-like term~$\Hill{1}{[S]}{K_{0.5}}$.

\section{Parameter estimation}\label{app:Parameters}
\subsection{Metabotropic receptors}
Rate constants~$O_{N},\,\Omega_{N}$~(\eqref{eq:GammaA}) lump information on astrocytic metabotropic receptors' activation and inactivation, namely how long it takes for these receptors, once bound by the agonist, to trigger \plcb-mediated \ip3\ production and how long this latter lasts. Since \ip3\ production mediated by agonist binding with the receptors controls the initial~intracellular \ca\ surge, these two rate constants may be estimated by rise times of agonist-triggered \ca\ signals. With this regard, experiments reported that application of~\SI{50}{\micro \Molar}~DHPG -- a potent agonist of mGluR5 which are the main type of metabotropic glutamate receptors expressed by astrocytes \citep{AronicaTroost2003} --, triggers submembrane~\ca\ signals characterized by a rise time~$\tau_{r}=0.272\pm \SI{0.095}{s}$. Because~mGluR5 affinity~($K_{0.5}$) for~DHPG is~$\sim \SI{2}{\micro \Molar}$ \citep{Brabet_NP1995}, that is much smaller than the applied agonist concentration, receptor saturation may be assumed in those experiments whereby the receptor activation rate by~DHPG~($O_{\textrm{DHPG}}$) can be expressed as a function of~$\tau_{r}$ \citep{Barbour_JN2001}, i.e.~$O_{\textrm{DHPG}}\approx \tau_{r}/(\SI{50}{\micro \Molar})=0.055-\SI{0.113}{\micro \Molar^{-1} s^{-1}}$, so that~$\Omega_{\textrm{DHPG}}=O_{\textrm{DHPG}}K_{0.5}\approx 0.11-\SI{0.22}{s^{-1}}$. Corresponding rate constants for glutamate may then be estimated assuming similar kinetics, yet with~$K_{0.5}=K_N=\Omega_N/O_N \approx 3-\SI{10}{\micro \Molar}$ \citep{Daggett_NP1995}, that is~1.5--5-fold larger than~$K_{0.5}$ for~DHPG. Moreover, since rise times of~\ca\ signals triggered by non-saturating physiological stimulation are faster than in the case of~DHPG \citep{Panatier_etal_Cell2011}, it may be assumed that~$O_N > O_\mathrm{DHPG}$. With this regard, for a choice of~$O_N \approx 3\times O_\textrm{DHPG}= \SI{0.3}{\micro \Molar^{-1} s^{-1}}$, with~$K_{N}=\SI{6}{\micro \Molar}$ such that~$\Omega_N = (\SI{0.3}{\micro \Molar^{-1}s^{-1}})(\SI{6}{\micro \Molar})=\SI{1.8}{s^{-1}}$, a peak of extracellular glutamate concentration of~$\SI{250}{\micro \Molar}$, delivered at $t=0$ and exponentially decaying at rate $\Omega_{c}=\SI{40}{s^{-1}}$ \citep{Clements1992}, is consistent with a peak fraction of bound receptors of~$\sim 0.75$ within~$\sim \SI{70}{\milli s}$ from stimulation (\eqref{eq:GammaA}), which is in good agreement with experimental rise times.

\subsection{\ip3R kinetics}
We consider a steady-state receptor open probability in the form of $p_\mathrm{open}(C,I) = \mathcal{H}_1^3(I,d_1)\cdot \cdot \mathcal{H}_1^3(C,d_5) (1-\Hill{1}{C}{Q_2})^3$ with $Q_2=d_2(I+d_1)/(I+d_3)$ (see \textcolor{red}{Chapter~3}) and choose parameters to fit corresponding experimental data by \citet{Ramos_BJ2000} for (i)~different \ca\ concentrations ($\hat{C}$ at a fixed \ip3\ level of $\bar{I}=\SI{1}{\micro \Molar}$, i.e. $\hat{p}(\hat{C})$; and (ii)~for different \ip3\ concentrations ($\hat{I}$) at an intracellular \ca\ concentration of $\bar{C}=\SI{25}{n \Molar}$, i.e. $\hat{p}(\hat{I})$. To reduce the problem dimensionality while retaining essential dynamical features of \ip3\ gating kinetics we set $d_1=d_3$ \citep{LiRinzel1994}. Accordingly, defining the vector parameter $\vec{x}_p = \rpar{d_1,d_2,d_5,O_2}$, we minimize the cost function $c_p(\vec{x}_p)=(p_\mathrm{open}(\hat{C},\bar{I})-\hat{p}(\hat{C}))^2 + (p_\mathrm{open}(\bar{C},\hat{I})-\hat{p}(\hat{I}))^2$ by the Artificial Bee Colony (ABC) algorithm \citep{Karaboga_JGO2007} considering 2000~evolutions of a colony of~100 individuals.

Ultrastructural analysis of astrocytes~\emph{in situ} revealed that the probability of~ER localization in the cytoplasmic space at the soma is between~$\sim$40--70\% \citep{Pivneva_CellCalcium2008}. This suggests that the corresponding ratio between ER and cytoplasmic volumes~($\rho_{A}$) is comprised between~$\sim$0.4--0.7. 

To estimate the cell's total free~\ca\ content~$C_{T}$ we make the consideration that the resting \ca\ concentration in the cytosol is $<\SI{0.15}{\micro \Molar}$ \citep{Zheng_Neuron2015} and can be neglected with respect to the amount of \ca\ stored in the ER ($C_{ER}$) \citep{Berridge_etal_NatRev2003}. Hence, with~$C_{ER}\ge \SI{10}{\micro \Molar}$ \citep{GovolinaScience1997} and a choice of~$\rho_{A}\ge 0.4$, it follows that $C_{T}\approx \rho_{A}\, C_{ER} \ge \SI{4}{\micro \Molar}$. In conditions close to store depletion during oscillations \citep{CamelloTepikin2002}, this latter value would also coincide with the peak~\ca\ reached in the cytoplasm, which is reported between $<\SI{5}{\micro \Molar}$ and $\sim \SI{20}{\micro \Molar}$ \citep{CsordasHajnoczky1999,ParpuraHaydon2000,KangOthmer_Chaos2009,Shigetomi_etal_Nature2010}. 

In our simulations we set $\rho_A = 0.5$ while leaving arbitrary the choice of $C_T$ as far as the resulting \ca\ oscillations qualitatively resemble the shape of those observed in experiments. The remaining parameters for CICR, i.e. $\vec{z}_c = \rpar{\Omega_C, O_P}$, were chosen to approximate the number and period of \ca\ oscillations observed \textit{on average} in experiments on cultured astrocytes that were stimulated by glutamate perfusion. By ``on average'' we mean that we considered the average trace resulting from $n=5$ different \ca\ signals generated within the same period of time and by the same stimulus in identical experimental conditions.

\subsection{\ip3 signaling}
Once set the CICR parameters, individual \ca\ traces used to obtained the above-mentioned ``average trace'' were used to search for $\vec{z}_p = \rpar{O_\beta,O_\delta,O_{3K},\Omega_{5P}}$, assuming random initial conditions. The ensuing parameter values were also used in Figures~\ref{fig:gchi-dynamics}--\ref{fig:gchidp} although $O_\beta,\,O_\delta$ and $O_{3K}$ were increased, from case to case, by a factor comprised between $1.2-2$ either to expand the oscillatory range or to promote CICR emergence (by increasing $O_\beta,\,O_\delta$) or termination (by larger $O_{3K}$ values).

\subsection{\ce{cPKC} and DAG signaling}
Calcium-dependent~cPKC-mediated phosphorylation has been documented for astrocytic\linebreak[4] mGluRs and~\ce{P_2Y_1Rs} \citep{CodazziTeruelMeyer2001,Hardy_Blood2005} and results in a reduction of receptor binding affinity by a factor~$\zeta \approx 2-10$ \citep{Hardy_Blood2005}, or possibly higher  depending on the cell's expression of~cPKCs \citep{Nakahara_JNC1997,Shinohara_PNAS2011}. Since experiments showed that~cPKC is robustly activated only when~\ca\ increases beyond half of the peak concentration reached during oscillations \citep{CodazziTeruelMeyer2001} then, considering peak \ca\ values of~$\sim 1-\SI{3}{\micro \Molar}$ \citep{Shigetomi_etal_Nature2010} allows estimating~\ca\ affinity of~cPKC in the range of~$K_{KC} \le 0.5-\SI{1.5}{\micro \Molar}$ which indeed comprises the value of $\sim \SI{700}{\nano \Molar}$ predicted experimentally \citep{Mosior_JBC1994}. Of the same order of magnitude also is the \ca\ affinity reported for DAGK, i.e. $K_{DC} \approx 0.3-\SI{0.4}{\micro \Molar}$ \citep{Sakane_JBC1991,Yamada_BJ1997}.

Reported values of DAG affinities for cPKC and DAGK may considerably differ. Micellar assays of cPKCs activity, suggests values of $K_{KD}$ as low as 4.6--\SI{13.3}{\nano \Molar} \citep{Ananthanarayanan_JBC2003}, whereas studies on purified DAGK suggest a substrate affinity for this kinase of $K_{DD}\approx \SI{60}{\micro \Molar}$ \citep{Kanoh_JBC1983}. The differences in experimental setups and the possibility that the activity of these kinases could be widely regulated by different DAG pools make these estimate of scarce utility for our model, where the DAG concentration is of the same order of magnitude of \ip3\ one. With this regard we choose to set these affinities to \SI{0.1}{\micro \Molar} which corresponded in our simulations to the average intracellular DAG concentration during \ca\ oscillations.

The remaining parameters, namely $\vec{z}_k = \rpar{O_{KD}, O_K, \Omega_D, O_D, \Omega_D}$ were arbitrarily chosen considering two constrains: (i)~DAG concentration for damped \ca\ oscillations must stabilize to a constant value; and (ii)~the down phase of \cpkc\ oscillations must follow that of \ca\ ones as suggested by experimental observations by \citet{CodazziTeruelMeyer2001}.

\section{Software}
The Python file \lstinline|figures.py| used to generate the figures of this chapter can be downloaded from the online book repository at \url{https://github.com/mdepitta/comp-glia-book}. The software for this chapter is organized in two folders. The \lstinline|data| folder contains data to fit the \gchi\ model. WebPlotDigitizer~4.0 (\url{https://automeris.io/WebPlotDigitizer}) was used to extract experimental data by \citet[][Figures~6 and~7]{Ramos_BJ2000} and \citet[][Figure~5]{CodazziTeruelMeyer2001}. The Jupyter notebook file \lstinline|data_loader.ipynb| found in this folder contains the code to load and clean experimental data used in the simulations. 

The \lstinline|code| folder contains instead all the routines (including \lstinline|figures.py|) used for the simulations of this chapter. The two files \lstinline|astrocyte_models.h| and \lstinline|astrocyte_models.cpp| contains the core \gchi\ model implementation in C/C++11, while the class \lstinline|Astrocyte| in\linebreak[4] \lstinline|astrocyte_models.py| provides the Python interface to simulate the \gchi\ model. The model was integrated by a variable-coefficient linear multistep Adams method in Nordsieck form which proved robust to correctly solve stiff problems rising from different parameter choices \citep{Skeel_MC1986}. Model fitting is provided by \lstinline|gchi_fit.py| and  relies on the PyGMO~2.6 optimization package (\url{https://github.com/esa/pagmo2.git}).

The library \lstinline|gchi_bifurcation.py| provides routines to estimate the period and range of \ca\ oscillation as in Figures~\ref{fig:gchidp}. These routines use numerical continuation of the extended \gchi\ model by the Python module PyDSTool~0.92 \citep[][\url{https://github.com/robclewley/pydstool}]{Clewley_PCB2012}.

\newpage
\section{Model parameters used in simulations}
\begin{table}[ht!]
\caption{Model parameters used in the simulations, unless differently specified in figure captions.}
\vspace{0.1cm}
\begin{tabularx}{\textwidth}[t]{l X l l}
    \hline
    Symbol      & Description                         &Value       &Units\\
    \hline
    \multicolumn{4}{c}{\textsl{Astrocyte receptors}}\\
    \hline
    $\Omega_N$  & Rate of receptor de-activation             & 1.8 &\si{s^{-1}}\\
    $O_N$       & Rate of agonist-mediated receptor activation & 0.3 &\si{\micro \Molar^{-1} s^{-1}}\\ 
    $n$         & Agonist binding cooperativity              & 1   &--\\        
    \hline
    \multicolumn{4}{c}{\textsl{\ip3R kinetics}}\\
    \hline
    $d_{1}$     & \ip3\ binding affinity              & 0.1        &\si{\micro \Molar}\\
    $O_{2}$     & Inactivating~\ca\ binding rate      & 0.325      &\si{\micro \Molar^{-1} s^{-1}}\\
    $d_{2}$     & Inactivating~\ca\ binding affinity  & 4.5		   &\si{\micro \Molar}\\
    $d_{3}$     & \ip3\ binding affinity~(with~\ca\ inactivation)  & 0.1  &\si{\micro \Molar}\\
    $d_{5}$     & Activating~\ca\ binding affinity    & 0.05       &\si{\micro \Molar}\\
    \hline
    \multicolumn{4}{c}{\textsl{\ca fluxes}}\\
    \hline
    $C_{T}$     & Total~ER~\ca\ content               & 5          &\si{\micro \Molar}\\
    $\rho_{A}$  & ER-to-cytoplasm volume ratio        & 0.5        &--\\
    $\Omega_{C}$& Maximal~\ca\ release rate by~\ip3Rs & 7.759      &\si{s^{-1}}\\
    $\Omega_{L}$& \ca\ leak rate                      & 0.1        &\si{s^{-1}}\\
    $O_{P}$     & Maximal~\ca\ uptake rate            & 5.499      &\si{\micro \Molar s^{-1}}\\
    $K_{P}$     & \ca\ affinity of~SERCA pumps        & 0.1        &\si{\micro \Molar}\\
    \hline
    \multicolumn{4}{c}{\textsl{\ip3\ production}}\\
	\hline    
    $O_{\beta}$ & Maximal rate of~\ip3\ production by~\plcb\ & 0.8   &\si{\micro \Molar s^{-1}}\\
    $O_{\delta}$& Maximal rate of~\ip3\ production by~\plcd\ & 0.025 &\si{\micro \Molar s^{-1}}\\
    $K_{\delta}$& \ca\ affinity of~\plcd\                    & 0.5   &\si{\micro \Molar}\\
    $\kappa_{\delta}$ & Inhibiting~\ip3\ affinity of~\plcd\  & 1.0   &\si{\micro \Molar}\\
    \hline
    \multicolumn{4}{c}{\textsl{\ip3\ degradation}}\\
	\hline    
    $\Omega_{5P}$ & Rate of~\ip3\ degradation by~\5p\         & 0.86 &\si{s^{-1}}\\
    $O_{3K}$      & Maximal rate of~\ip3\ degradation by~\3k\ & 0.86 &\si{\micro \Molar s^{-1}}\\
    $K_{D}$       & \ca\ affinity of~\3k\                     & 0.5 &\si{\micro \Molar}\\
    $K_{3K}$      & \ip3\ affinity of~\3k\                    & 1.0 &\si{\micro \Molar}\\
    \hline
    \multicolumn{4}{c}{\textsl{\ce{DAG} dynamics}}\\
	\hline    
    $\Omega_D$  & Unspecific rate of degradation              & 0.26 &\si{s^{-1}}\\
    $O_D$       & Rate of degradation by DAGK                 & 0.45 &\si{\micro \Molar s^{-1}}\\
    $K_{DC}$    & DAGK affinity for \ca\                      & 0.3 &\si{\micro \Molar}\\
    $K_{DD}$    & DAGK affinity for DAG                       & 0.1 &\si{\micro \Molar}\\
    \hline
    \multicolumn{4}{c}{\textsl{\ce{cPKC} signaling}}\\
	\hline        
	$O_{KD}$    & Rate of \cpkc\ production                  & 0.28 &\si{\micro \Molar s^{-1}}\\
	$\Omega_{KD}$ & Rate of \cpkc\ deactivation              & 0.33 &\si{s^{-1}}\\
    $K_{KC}$    & \ca\ affinity of PKC                       & 0.5  &\si{\micro \Molar}\\
    $O_{K}$     & Rate of receptor phosphorylation           & 1.0  &\si{\micro \Molar^{-1} s^{-1}}\\
    \hline\\
\end{tabularx}
\label{tab:Model-Parameters}
\end{table}
\end{appendices}

\newpage
\clearpage

\newpage
\clearpage

\newpage
\clearpage

\newpage
\clearpage

\newpage
\clearpage

\newpage
\clearpage

\newpage
\bibliography{./ch5_depitta.bib}

\begin{thebibliography}{}

\bibitem[Allen and Barres, 2009]{AllenBarres_Nature2009}
Allen, N.~J. and Barres, B.~A. (2009).
\newblock Glia - more than just brain glue.
\newblock {\em Nature}, 457:675--677.

\bibitem[Ananthanarayanan et~al., 2003]{Ananthanarayanan_JBC2003}
Ananthanarayanan, B., Stahelin, R.~V., Digman, M.~A., and Cho, W. (2003).
\newblock Activation mechanisms of conventional protein kinase~{C} isoforms are
  determined by the ligand affinity and conformational flexibility of
  their~{C1} domains.
\newblock {\em Journal of Biological Chemistry}, 278(47):46886--46894.

\bibitem[Aronica et~al., 2003]{AronicaTroost2003}
Aronica, E., Gorter, J.~A., {Ijlst-Keizers}, H., Rozemuller, A.~J., Yankaya,
  B., Leenstra, S., and Troost, D. (2003).
\newblock Expression and functional role of {mGluR3} and {mGluR5} in human
  astrocytes and glioma cells: opposite regulation of glutamate transporter
  proteins.
\newblock {\em Eur. J. Neurosci.}, 17:2106--2118.

\bibitem[Barbour, 2001]{Barbour_JN2001}
Barbour, B. (2001).
\newblock An evaluation of synapse independence.
\newblock {\em J. Neurosci.}, 21(20):7969--7984.

\bibitem[Bekar et~al., 2008]{Bekar_CC2008}
Bekar, L.~K., He, W., and Nedergaard, M. (2008).
\newblock Locus coeruleus $\alpha$-adrenergic--mediated activation of cortical
  astrocytes in vivo.
\newblock {\em Cerebral cortex}, 18(12):2789--2795.

\bibitem[Berridge et~al., 2003]{Berridge_etal_NatRev2003}
Berridge, M.~J., Bootman, M.~D., and Roderick, H.~L. (2003).
\newblock Calcium signalling: dynamics, homeostasis and remodelling.
\newblock {\em Nat. Rev. Mol. Cell. Biol.}, 4:517--529.

\bibitem[Berridge and Irvine, 1989]{BerridgeIrvine_Nature1989}
Berridge, M.~J. and Irvine, R.~F. (1989).
\newblock Inositol phosphates and cell signalling.
\newblock {\em Nature}, 341(6239):197--205.

\bibitem[Bindocci et~al., 2017]{Bindocci_Science2017}
Bindocci, E., Savtchouk, I., Liaudet, N., Becker, D., and Carriero,
  G.and~Volterra, A. (2017).
\newblock Three-dimensional {Ca$^{2+}$} imaging advances understanding of
  astrocyte biology.
\newblock {\em Science}, 356:6339.

\bibitem[Brabet et~al., 1995]{Brabet_NP1995}
Brabet, I., Mary, S., Bockaert, J., and Pin, J. (1995).
\newblock Phenylglycine derivatives discriminate between {mGluR1- and
  mGluR5-mediated} responses.
\newblock {\em Neuropharmacology}, 34(8):895--903.

\bibitem[Brambilla et~al., 1999]{Brambilla_BJP1999}
Brambilla, R., Burnstock, G., Bonazzi, A., Ceruti, S., Cattabeni, F., and
  Abbracchio, M.~P. (1999).
\newblock Cyclo-oxygenase-2 mediates~{P2Y} receptor-induced reactive
  astrogliosis.
\newblock {\em British Journal of Pharmacology}, 126(3):563--567.

\bibitem[Bruner and Murphy, 1990]{BrunerMurphy_JNC1990}
Bruner, G. and Murphy, S. (1990).
\newblock {ATP-evoked arachidonic acid mobilization in astrocytes is via a
  P2Y-purinergic receptor}.
\newblock {\em Journal of Neurochemistry}, 55(5):1569--1575.

\bibitem[Camello et~al., 2002]{CamelloTepikin2002}
Camello, C., Lomax, R., Petersen, O.~H., and Tepikin, A.~V. (2002).
\newblock Calcium leak from intracellular stores - the enigma of calcium
  signalling.
\newblock {\em Cell Calcium}, 32(5-6):355--361.

\bibitem[Carmignoto, 2000]{Carmignoto2000}
Carmignoto, G. (2000).
\newblock Reciprocal communication systems between astrocytes and neurones.
\newblock {\em Prog. Neurobiol.}, 62:561--581.

\bibitem[Carrasco and M{\'e}rida, 2007]{Carrasco_TiBS2007}
Carrasco, S. and M{\'e}rida, I. (2007).
\newblock Diacylglycerol, when simplicity becomes complex.
\newblock {\em Trends in Biochemical Sciences}, 32(1):27--36.

\bibitem[Changeux and Edelstein, 2005]{Changeux_Science2005}
Changeux, J. and Edelstein, S.~J. (2005).
\newblock Allosteric mechanisms of signal transduction.
\newblock {\em Science}, 308(5727):1424--1428.

\bibitem[Chen et~al., 2012]{Chen_PNAS2012}
Chen, N., Sugihara, H., Sharma, J., Perea, G., Petravicz, J., Le, C., and Sur,
  M. (2012).
\newblock Nucleus basalis enabled stimulus specific plasticity in the visual
  cortex is mediated by astrocytes.
\newblock {\em Proc. Natl. Acad. Sci. USA}, 109(41):E2832�E2841.

\bibitem[Clements et~al., 1992]{Clements1992}
Clements, J.~D., Lester, R. A.~J., Tong, G., Jahr, C.~E., and Westbrook, G.~L.
  (1992).
\newblock The time course of glutamate in the synaptic cleft.
\newblock {\em Science}, 258:1498--1501.

\bibitem[Clewley, 2012]{Clewley_PCB2012}
Clewley, R. (2012).
\newblock Hybrid models and biological model reduction with{PyDSTool}.
\newblock {\em PLoS Computational Biology}, 8(8):e1002628.

\bibitem[Codazzi et~al., 2001]{CodazziTeruelMeyer2001}
Codazzi, F., Teruel, M.~N., and Meyer, T. (2001).
\newblock Control of astrocyte {Ca$^{2+}$} oscillations and waves by
  oscillating translocation and activation of protein kinase {C}.
\newblock {\em Curr. Biol.}, 11(14):1089--1097.

\bibitem[Communi et~al., 1999]{CommuniErneux_JBC1999}
Communi, D., Dewaste, V., and Erneux, C. (1999).
\newblock Calcium-calmodulin-dependent protein kinase~{II} and protein
  kinase~{C}-mediated phosphorylation and activation of {D}-myo-inositol
  1,4,5-trisphosphate 3-kinase~{B} in astrocytes.
\newblock {\em J. Biol. Chem.}, 274:14734--14742.

\bibitem[Communi et~al., 2001]{CommuniErneux2001}
Communi, D., Gevaert, K., Demol, H., Vandekerckhove, J., and Erneux, C. (2001).
\newblock A novel receptor-mediated regulation mechanism of type~{I} inositol
  polyphosphate 5-phosphatase by calcium/calmodulin-dependent protein
  kinase~{II} phosphorylation.
\newblock {\em J. Biol. Chem.}, 276(42):38738--38747.

\bibitem[Communi et~al., 1995]{CommuniErneux_Rev1995}
Communi, D., Vanweyenberg, V., and Erneux, C. (1995).
\newblock Molecular study and regulation of {D}-myo-inositol
  1,4,5-trisphopshate 3-kinase.
\newblock {\em Cell. Signal.}, 7(7):643--650.

\bibitem[Communi et~al., 1997]{CommuniErneux1997}
Communi, D., Vanweyenberg, V., and Erneux, C. (1997).
\newblock \textsc{d}-\textit{myo}-inositol 1,4,5-trisphosphate 3-kinase~{A} is
  activated by receptor activation through a calcium:calmodulin-dependent
  protein kinase~{II} phosphorylation mechanism.
\newblock {\em EMBO J.}, 16(8):1943--1952.

\bibitem[Connolly et~al., 1987]{Connolly_JBC1987}
Connolly, T., Bansal, V., Bross, T., Irvine, R., and Majerus, P. (1987).
\newblock The metabolism of tris-and tetraphosphates of inositol by
  5-phosphomonoesterase and 3-kinase enzymes.
\newblock {\em J. Biol. Chem.}, 262(5):2146--2149.

\bibitem[Crist{\'o}v{\~a}o-Ferreira et~al., 2013]{Cristovao-Ferreira_PS2013}
Crist{\'o}v{\~a}o-Ferreira, S., Navarro, G., Brugarolas, M., P{\'e}rez-Capote,
  K., Vaz, S.~H., Fattorini, G., Conti, F., Lluis, C., Ribeiro, J.~A.,
  {McCormick}, P.~J., Casad\'o, V., Franco, R., and Sebasti{\~a}o, A.~M.
  (2013).
\newblock {A$_1$R--A$_{2A}$R} heteromers coupled to g$_s$ and g$_{i/0}$
  proteins modulate {GABA} transport into astrocytes.
\newblock {\em Purinergic signalling}, 9(3):433--449.

\bibitem[{Csord\`as} et~al., 1999]{CsordasHajnoczky1999}
{Csord\`as}, G., Thomas, A.~P., and {Hajn\'oczky}, G. (1999).
\newblock Quasi-synaptic calcium signal transmission between endoplasmic
  reticulum and mitochondria.
\newblock {\em EMBO J.}, 18(1):96--108.

\bibitem[Cui et~al., 2016]{Cui_eLife2016}
Cui, Y., Prokin, I., Xu, H., Delord, B., Genet, S., Venance, L., and Berry, H.
  (2016).
\newblock Endocannabinoid dynamics gate spike-timing dependent depression and
  potentiation.
\newblock {\em Elife}, 5:e13185.

\bibitem[Daggett et~al., 1995]{Daggett_NP1995}
Daggett, L., Sacaan, A., Akong, M., Rao, S., Hess, S., Liaw, C., Urrutia, A.,
  Jachec, C., Ellis, S., Dreessen, J., et~al. (1995).
\newblock Molecular and functional characterization of recombinant human
  metabotropic glutamate receptor subtype~5.
\newblock {\em Neuropharmacology}, 34(8):871--886.

\bibitem[{De Konick} and Schulman, 1998]{DeKonickSchulman1998}
{De Konick}, P. and Schulman, H. (1998).
\newblock Sensitivity of {CaM} kinase~{II} to the frequency of {Ca$^{2+}$}
  oscillations.
\newblock {\em Science}, 279:227--230.

\bibitem[{De Pitt\`a} et~al., 2009]{DePitta_JOBP2009}
{De Pitt\`a}, M., Goldberg, M., Volman, V., Berry, H., and Ben-Jacob, E.
  (2009).
\newblock Glutamate-dependent intracellular calcium and {IP$_{3}$} oscillating
  and pulsating dynamics in astrocytes.
\newblock {\em J. Biol. Phys.}, 35:383--411.

\bibitem[{De Pitt\`a} et~al., 2013]{DePitta_FCN13}
{De Pitt\`a}, M., Volman, V., Berry, H., Parpura, V., Liaudet, N., Volterra,
  A., and Ben-Jacob, E. (2013).
\newblock Computational quest for understanding the role of astrocyte signaling
  in synaptic transmission and plasticity.
\newblock {\em Front. Comp. Neurosci.}, 6:98.

\bibitem[Di~Castro et~al., 2011]{DiCastro_Volterra_NatNeurosci2011}
Di~Castro, M., Chuquet, J., Liaudet, N., Bhaukaurally, K., Santello, M.,
  Bouvier, D., Tiret, P., and Volterra, A. (2011).
\newblock Local {Ca$^{2+}$} detection and modulation of synaptic release by
  astrocytes.
\newblock {\em Nat. Neurosci.}, 14:1276�1284.

\bibitem[Ding et~al., 2013]{Ding_CC2013}
Ding, F., O’Donnell, J., Thrane, A.~S., Zeppenfeld, D., Kang, H., Xie, L.,
  Wang, F., and Nedergaard, M. (2013).
\newblock {$\alpha_1$-Adrenergic} receptors mediate coordinated {Ca$^{2+}$}
  signaling of cortical astrocytes in awake, behaving mice.
\newblock {\em Cell Calcium}, 54(6):387--394.

\bibitem[Doengi et~al., 2009]{Doengi_PNAS2009}
Doengi, M., Hirnet, D., Coulon, P., Pape, H.-C., Deitmer, J.~W., and Lohr, C.
  (2009).
\newblock {GABA}~uptake-dependent {Ca$^{2+}$} signaling in developing olfactory
  bulb astrocytes.
\newblock {\em Proceedings of the National Academy of Sciences},
  106(41):17570--17575.

\bibitem[Dominguez et~al., 2013]{Dominguez_CD2013}
Dominguez, C.~L., Floyd, D.~H., Xiao, A., Mullins, G.~R., Kefas, B.~A., Xin,
  W., Yacur, M.~N., Abounader, R., Lee, J.~K., Wilson, G.~M., Harris, T.~E.,
  and Purow, B.~W. (2013).
\newblock Diacylglycerol kinase $\alpha$ is a critical signaling node and novel
  therapeutic target in glioblastoma and other cancers.
\newblock {\em Cancer Discovery}, 3(7):782--797.

\bibitem[Dupont and Erneux, 1997]{DupontErneux1997}
Dupont, G. and Erneux, C. (1997).
\newblock Simulations of the effects of inositol 1,4,5-trisphosphate 3-kinase
  and 5-phosphatase activities on {Ca$^{2+}$} oscillations.
\newblock {\em Cell Calcium}, 22(5):321--331.

\bibitem[Erneux et~al., 1998]{Erneux_BBA1998}
Erneux, C., Govaerts, C., Communi, D., and Pesesse, X. (1998).
\newblock The diversity and possible functions of the inositol polyphosphate
  5-phosphatases.
\newblock {\em Biochim. Biophys. Acta}, 1436(1-2):185--189.

\bibitem[Essen et~al., 1996]{EssenWilliams_Nature1996}
Essen, L., Perisic, O., Cheung, R., Katan, M., and Williams, R.~L. (1996).
\newblock Crystal structure of a mammalian phosphoinositide-specific
  phospholipase~{C}.
\newblock {\em Nature}, 380:595--602.

\bibitem[Essen et~al., 1997]{EssenWilliams_Biochem1997}
Essen, L., Perisic, O., Lunch, D.~E., Katan, M., and Williams, R.~L. (1997).
\newblock A ternary metal binding site in the {C2} domain of
  phosphoinositide-specific phospholipase~{C-$\delta$1}.
\newblock {\em Biochemistry (Mosc.)}, 37(10):4568--4680.

\bibitem[Fam et~al., 2000]{Fam_JN2000}
Fam, S., Gallagher, C., and Salter, M. (2000).
\newblock {P2Y$_{1}$} purinoceptor-mediated~{Ca$^{2+}$} signaling
  and~{Ca$^{2+}$} wave propagation in dorsal spinal cord astrocytes.
\newblock {\em J. Neurosci.}, 20(8):2800--2808.

\bibitem[Fisher, 1995]{Fisher_RevEJP1995}
Fisher, S.~K. (1995).
\newblock Homologous and heterologous regulation of receptor-stimulated
  phosphoinosited hydrolysis.
\newblock {\em Eur. J. Pharmacol.}, 288:231--250.

\bibitem[Gallo and Ghiani, 2000]{GalloGhiani2000}
Gallo, V. and Ghiani, A. (2000).
\newblock Glutamate receptors in glia: new cells, new inputs and new functions.
\newblock {\em Trends Pharm. Sci.}, 21:252--258.

\bibitem[Giaume et~al., 1991]{Giaume_PNAS1991}
Giaume, C., Marin, P., Cordier, J., Glowinski, J., and Premont, J. (1991).
\newblock Adrenergic regulation of intercellular communications between
  cultured striatal astrocytes from the mouse.
\newblock {\em Proceedings of the National Academy of Sciences},
  88(13):5577--5581.

\bibitem[Golovina and Blaustein, 1997]{GovolinaScience1997}
Golovina, V.~A. and Blaustein, M.~P. (1997).
\newblock Spatially and functionally distinct~{Ca$^{2+}$} stores in
  sarcoplasmic and endoplasmic reticulum.
\newblock {\em Science}, 275:1643--1648.

\bibitem[Griner and Kazanietz, 2007]{Griner_NRC2007}
Griner, E.~M. and Kazanietz, M.~G. (2007).
\newblock Protein kinase~{C} and other diacylglycerol effectors in cancer.
\newblock {\em Nature Reviews Cancer}, 7(4).

\bibitem[Hanson et~al., 1994]{HansonSchulman_Neuron1994}
Hanson, P.~I., Meyer, T., Stryer, L., and Schulman, H. (1994).
\newblock Dual role of calmodulin in autophosphorylation of multifunctional
  {CaM} kinase may underlie decoding of calcium signals.
\newblock {\em Neuron}, 12:943--956.

\bibitem[Hardy et~al., 2005]{Hardy_Blood2005}
Hardy, A., Conley, P., Luo, J., Benovic, J., Poole, A., and Mundell, S. (2005).
\newblock {P2Y1~and~P2Y12} receptors for~{ADP} desensitize by distinct
  kinase-dependent mechanisms.
\newblock {\em Blood}, 105(9):3552--3560.

\bibitem[Heller and Rusakov, 2015]{Heller_Glia2015}
Heller, J.~P. and Rusakov, D.~A. (2015).
\newblock Morphological plasticity of astroglia: understanding synaptic
  microenvironment.
\newblock {\em Glia}, 63(12):2133--2151.

\bibitem[Hermosura et~al., 2000]{Hermosura_Nature2000}
Hermosura, M., Takeuchi, H., Fleig, A., Riley, A., Potter, B., Hirata, M., and
  Penner, R. (2000).
\newblock {InsP4 facilitates store-operated calcium influx by inhibition of
  InsP3~5-phosphatase}.
\newblock {\em Nature}, 408(6813):735--740.

\bibitem[H\"ofer et~al., 2002]{HoferGiaume2002}
H\"ofer, T., Venance, L., and Giaume, C. (2002).
\newblock Control and plasticity of intercellular calcium waves in astrocytes:
  a modeling approach.
\newblock {\em J. Neurosci.}, 22(12):4850--4859.

\bibitem[Irvine et~al., 1986]{IrvineBerridge_Nat1986}
Irvine, R.~F., Letcher, A.~J., Heslop, J.~P., and Berridge, M.~J. (1986).
\newblock The inositol tris/tetrakisphopshate pathway--demonstration of
  {Ins(1,4,5)P$_{3}$} 3-kinase activity in animal tissues.
\newblock {\em Nature}, 320:631--634.

\bibitem[Irvine and Schell, 2001]{Irvine_NRMCB2001}
Irvine, R.~F. and Schell, M.~J. (2001).
\newblock Back in the water: the return of the inositol phosphates.
\newblock {\em Nat. Rev. Mol. Cell Biol.}, 2(5):327--338.

\bibitem[Jennings et~al., 2017]{Jennings_Glia2017}
Jennings, A., Tyurikova, O., Bard, L., Zheng, K., Semyanov, A., Henneberger,
  C., and Rusakov, D.~A. (2017).
\newblock Dopamine elevates and lowers astroglial~{Ca$^{2+}$} through distinct
  pathways depending on local synaptic circuitry.
\newblock {\em Glia}, 65(3):447--459.

\bibitem[Jourdain et~al., 2007]{JourdainVolterra2007}
Jourdain, P., Bergersen, L.~H., Bhaukaurally, K., Bezzi, P., Santello, M.,
  Domercq, M., Matute, C., Tonello, F., Gundersen, V., and Volterra, A. (2007).
\newblock Glutamate exocytosis from astrocytes controls synaptic strength.
\newblock {\em Nat. Neurosci.}, 10(3):331--339.

\bibitem[Kang et~al., 1998]{KangNedergaard1998}
Kang, J., Jiang, L., Goldman, S.~A., and Nedergaard, M. (1998).
\newblock Astrocyte-mediated potentiation of inhibitory synaptic transmission.
\newblock {\em Nat. Neurosci.}, 1(8):683--692.

\bibitem[Kang and Othmer, 2009]{KangOthmer_Chaos2009}
Kang, M. and Othmer, H. (2009).
\newblock Spatiotemporal characteristics of calcium dynamics in astrocytes.
\newblock {\em Chaos}, 19(3):037116.

\bibitem[Kanoh et~al., 1983]{Kanoh_JBC1983}
Kanoh, H., Kondoh, H., and Ono, T. (1983).
\newblock {Diacylglycerol kinase from pig brain. Purification and phospholipid
  dependencies}.
\newblock {\em Journal of Biological Chemistry}, 258(3):1767--1774.

\bibitem[Karaboga and Basturk, 2007]{Karaboga_JGO2007}
Karaboga, D. and Basturk, B. (2007).
\newblock A powerful and efficient algorithm for numerical function
  optimization: artificial bee colony~{(ABC)} algorithm.
\newblock {\em Journal of Global Optimization}, 39(3):459--471.

\bibitem[Keener and Sneyd, 2008]{KeenerSneyd_2008_Book}
Keener, J. and Sneyd, J. (2008).
\newblock {\em Mathematical Physiology:~I:~Cellular Physiology}, volume~1.
\newblock Springer.

\bibitem[Kolodziej et~al., 2000]{KolodziejStoops_JBC2000}
Kolodziej, S.~J., Hudmon, A., Waxham, M.~N., and Stoops, J.~K. (2000).
\newblock Three-dimensional reconstructions of calcium/calmodulin-dependent
  {(CaM)} kinase~{II}$\alpha$ and truncated {CaM kinase II}$\alpha$ reveal a
  unique organization for its structural core and functional domains.
\newblock {\em J. Biol. Chem.}, 275(19):14354--14359.

\bibitem[Lemon et~al., 2003]{Lemon_JTB2003}
Lemon, G., Gibson, W.~G., and Bennett, M.~R. (2003).
\newblock Metabotropic receptor activation, desensitization and sequestration
  -- {I}: modelling calcium and inositol 1,4,5-trisphosphate dynamics following
  receptor activation.
\newblock {\em Journal of Theoretical Biology}, 223(1):93--111.

\bibitem[Li and Rinzel, 1994]{LiRinzel1994}
Li, Y. and Rinzel, J. (1994).
\newblock Equations for {InsP$_{3}$} receptor-mediated {[Ca$^{2+}$]$_{i}$}
  oscillations derived from a detailed kinetic model: {A Hodgkin-Huxley} like
  formalism.
\newblock {\em J. Theor. Biol.}, 166:461--473.

\bibitem[Losi et~al., 2014]{Losi_PTRSB2014}
Losi, G., Mariotti, L., and Carmignoto, G. (2014).
\newblock {GABAergic} interneuron to astrocyte signalling: a neglected form of
  cell communication in the brain.
\newblock {\em Phil. Trans. Royal Soc. B: Biological Sciences},
  369(1654):20130609.

\bibitem[Marinissen and Gutkind, 2001]{Marinissen_TiPS2001}
Marinissen, M.~J. and Gutkind, J.~S. (2001).
\newblock {G-protein-coupled} receptors and signaling networks: emerging
  paradigms.
\newblock {\em Trends in Pharmacological Sciences}, 22(7):368--376.

\bibitem[Mariotti et~al., 2016]{Mariotti_Glia2016}
Mariotti, L., Losi, G., Sessolo, M., Marcon, I., and Carmignoto, G. (2016).
\newblock The inhibitory neurotransmitter~{GABA} evokes
  long-lasting~{Ca$^{2+}$} oscillations in cortical astrocytes.
\newblock {\em Glia}, 64(3):363--373.

\bibitem[Mart{\'i}n et~al., 2015]{Martin_Science2015}
Mart{\'i}n, R., {Bajo-Gra{\~n}eras}, R., Moratalla, R., Perea, G., and Araque,
  A. (2015).
\newblock Circuit-specific signaling in astrocyte-neuron networks in basal
  ganglia pathways.
\newblock {\em Science}, 349(6249):730--734.

\bibitem[M{\'e}rida et~al., 2008]{Merida_BJ2008}
M{\'e}rida, I., {\'A}vila-Flores, A., and Merino, E. (2008).
\newblock Diacylglycerol kinases: at the hub of cell signalling.
\newblock {\em Biochemical Journal}, 409(1):1--18.

\bibitem[Min and Nevian, 2012]{MinNevian_NatNeurosci2012}
Min, R. and Nevian, T. (2012).
\newblock Astrocyte signaling controls spike timing-dependent depression at
  neocortical synapses.
\newblock {\em Nat. Neurosci.}, 15(5):746--753.

\bibitem[Mishra and Bhalla, 2002]{MishraBhalla2002}
Mishra, J. and Bhalla, U.~S. (2002).
\newblock Simulations of inositol phosphate metabolism and its interaction with
  \textsc{I}ns\textsc{P}$_{3}$-mediated calcium release.
\newblock {\em Biophys. J.}, 83:1298--1316.

\bibitem[Mosior and Epand, 1994]{Mosior_JBC1994}
Mosior, M. and Epand, R.~M. (1994).
\newblock Characterization of the calcium-binding site that regulates
  association of protein kinase~{C} with phospholipid bilayers.
\newblock {\em Journal of Biological Chemistry}, 269(19):13798--13805.

\bibitem[Nakahara et~al., 1997]{Nakahara_JNC1997}
Nakahara, K., Okada, M., and Nakanishi, S. (1997).
\newblock The metabotropic glutamate receptor~{mGluR5} induces calcium
  oscillations in cultured astrocytes via protein {kinase~C} phosphorylation.
\newblock {\em J. Neurochem.}, 69(4):1467--1475.

\bibitem[Navarrete and Araque, 2008]{NavarreteAraque_Neuron2008}
Navarrete, M. and Araque, A. (2008).
\newblock Endocannabinoids mediate neuron-astrocyte communication.
\newblock {\em Neuron}, 57(6):883--893.

\bibitem[Navarrete et~al., 2012]{Navarrete_PB2012}
Navarrete, M., Perea, G., de~Sevilla, D., G{\'o}mez-Gonzalo, M., N{\'u}{\~n}ez,
  A., Mart{\'\i}n, E., and Araque, A. (2012).
\newblock Astrocytes mediate in vivo cholinergic-induced synaptic plasticity.
\newblock {\em PLoS Biol.}, 10(2):e1001259.

\bibitem[Nishizuka, 1995]{NishizukaRev1995}
Nishizuka, Y. (1995).
\newblock Protein kinase~{C} and lipid signaling for sustained cellular
  responses.
\newblock {\em FASEB J.}, 9:484--496.

\bibitem[Oancea and Meyer, 1998]{Oancea_Cell1998}
Oancea, E. and Meyer, T. (1998).
\newblock Protein kinase~{C} as a molecular machine for decoding calcium and
  diacylglycerol signals.
\newblock {\em Cell}, 95:307--318.

\bibitem[Ochocka and Pawelczyk, 2003]{Ochocka_ABP2003}
Ochocka, A.-M. and Pawelczyk, T. (2003).
\newblock Isozymes delta of phosphoinositide-specific phospholipase~{C} and
  their role in signal transduction in the cell.
\newblock {\em Acta Biochimica Polonica}, 50(4):1097--1110.

\bibitem[Overington et~al., 2006]{Overington_NRDD2006}
Overington, J.~P., {Al-Lazikani}, B., and Hopkins, A.~L. (2006).
\newblock How many drug targets are there?
\newblock {\em Nature Reviews Drug discovery}, 5(12):993--996.

\bibitem[Panatier et~al., 2011]{Panatier_etal_Cell2011}
Panatier, A., Vall{\'e}e, J., Haber, M., Murai, K., Lacaille, J., and
  Robitaille, R. (2011).
\newblock Astrocytes are endogenous regulators of basal transmission at central
  synapses.
\newblock {\em Cell}, 146:785--798.

\bibitem[Parpura and Haydon, 2000]{ParpuraHaydon2000}
Parpura, V. and Haydon, P.~G. (2000).
\newblock Physiological astrocytic calcium levels stimulate glutamate release
  to modulate adjacent neurons.
\newblock {\em Proc. Natl. Acad. Sci. USA}, 97(15):8629--8634.

\bibitem[Parri and Crunelli, 2003]{ParriCrunelli_Neuroscience2003}
Parri, H.~R. and Crunelli, V. (2003).
\newblock The role of~{Ca$^{2+}$} in the generation of spontaneous
  astrocytic~{Ca$^{2+}$} oscillations.
\newblock {\em Neuroscience}, 120(4):979--992.

\bibitem[Pawelczyk and Matecki, 1997]{PawelczykMatecki_EurJBiochem1997}
Pawelczyk, T. and Matecki, A. (1997).
\newblock Structural requirements of phospholipase~{C}~$\delta 1$ for
  regulation by spermine, sphingosine and sphingomyelin.
\newblock {\em Eur. J. Biochem.}, 248:459--465.

\bibitem[Perea and Araque, 005a]{PereaAraque_JNT2005}
Perea, G. and Araque, A. (2005a).
\newblock Synaptic regulation of the astrocyte calcium signal.
\newblock {\em J. Neur. Transmission}, 112:127--135.

\bibitem[Pivneva et~al., 2008]{Pivneva_CellCalcium2008}
Pivneva, T., Haas, B., Reyes-Haro, D., Laube, G., Veh, R., Nolte, C., Skibo,
  G., and Kettenmann, H. (2008).
\newblock Store-operated {Ca$^{2+}$} entry in astrocytes: different spatial
  arrangement of endoplasmic reticulum explains functional diversity in vitro
  and in situ.
\newblock {\em Cell Calcium}, 43(6):591--601.

\bibitem[{Ramos-Franco} et~al., 2000]{Ramos_BJ2000}
{Ramos-Franco}, J., Bare, D., Caenepeel, S., Nani, A., Fill, M., and Mignery,
  G. (2000).
\newblock Single-channel function of recombinant type~2 inositol
  1,4,5-trisphosphate receptor.
\newblock {\em Biophys. J.}, 79(3):1388--1399.

\bibitem[Rebecchi and Pentyala, 2000]{RebecchiPentyala2000}
Rebecchi, M.~J. and Pentyala, S.~N. (2000).
\newblock Structure, function, and control of phosphoinositide-specific
  phospholipase {C}.
\newblock {\em Physiol. Rev.}, 80(4):1291--1335.

\bibitem[Rhee and Bae, 1997]{RheeBaeRev1997}
Rhee, S.~G. and Bae, Y.~S. (1997).
\newblock Regulation of phosphoinositide-specific phospholipase~{C} isozymes.
\newblock {\em J. Biol. Chem.}, 272:15045--15048.

\bibitem[Rosenberger et~al., 2007]{Rosenberger_Lipids2007}
Rosenberger, T.~A., Farooqui, A.~A., and Horrocks, L.~A. (2007).
\newblock Bovine brain diacylglycerol lipase: substrate specificity and
  activation by cyclic{AMP}-dependent protein kinase.
\newblock {\em Lipids}, 42(3):187--195.

\bibitem[Ryu et~al., 1990]{RyuRhee_JBC1990}
Ryu, S.~H., Kin, U., Wahl, M.~I., Brown, a.~b., Carpenter, G., Huang, K., and
  Rhee, S.~G. (1990).
\newblock Feedback regulation of phospholipase {C}-$\beta$ by protein
  kinase~{C}.
\newblock {\em J. Biol. Chem.}, 265(29):17941--17945.

\bibitem[Sakane et~al., 1991]{Sakane_JBC1991}
Sakane, F., Yamada, K., Imai, S.-I., and Kanoh, H. (1991).
\newblock Porcine~{80-kDa} diacylglycerol kinase is a calcium-binding and
  calcium/phospholipid-dependent enzyme and undergoes calcium-dependent
  translocation.
\newblock {\em Journal of Biological Chemistry}, 266(11):7096--7100.

\bibitem[Santello and Volterra, 2012]{Santello_TiNS2012}
Santello, M. and Volterra, A. (2012).
\newblock {TNF$\alpha$}~in synaptic function: switching gears.
\newblock {\em Trends in Neurosci.}, 35(10):638--647.

\bibitem[Serrano et~al., 2006]{Serrano_etal_JN2006}
Serrano, A., Haddjeri, N., Lacaille, J., and Robitaille, R. (2006).
\newblock {GABAergic} network activation of glial cells underlies
  heterosynaptic depression.
\newblock {\em J. Neurosci.}, 26(20):5370--5382.

\bibitem[Shigetomi et~al., 2010]{Shigetomi_etal_Nature2010}
Shigetomi, E., Kracun, S., Sovfroniew, M.~S., and Khakh, B.~S. (2010).
\newblock A genetically targeted optical sensor to monitor calcium signals in
  astrocyte processes.
\newblock {\em Nat. Neurosci.}, 13(6):759--766.

\bibitem[Shinohara et~al., 2011]{Shinohara_PNAS2011}
Shinohara, T., Michikawa, T., Enomoto, M., Goto, J., Iwai, M., Matsu-ura, T.,
  Yamazaki, H., Miyamoto, A., Suzuki, A., and Mikoshiba, K. (2011).
\newblock Mechanistic basis of bell-shaped dependence of inositol
  1,4,5-trisphosphate receptor gating on cytosolic calcium.
\newblock {\em Proc. Natl. Acad. Sci. USA}, 108(37):15486--15491.

\bibitem[Shinomura et~al., 1991]{ShinomuraNishizuka1991}
Shinomura, T., Asaoka, Y., Oka, M., Yoshida, K., and Nishizuka, Y. (1991).
\newblock Synergistic action of diacylglycerol and unsaturated fatty acid for
  protein kinase~{C} activation: {Its} possible implications.
\newblock {\em Proc. Natl. Acad. Sci. USA}, 88:5149--5153.

\bibitem[Sim et~al., 1990]{SimRhee_JBC1990}
Sim, S.~S., Kim, J.~W., and Rhee, S.~G. (1990).
\newblock Regulation of {D-myo-inositol} 1,4,5-trisphosphate 3-kinase by
  {cAMP}-dependent protein kinase and protein kinase~c.
\newblock {\em J. Biol. Chem.}, 265:10367--10372.

\bibitem[Sims and Allbritton, 1998]{SimsAllbritton1998}
Sims, C.~E. and Allbritton, N.~L. (1998).
\newblock Metabolism of inositol 1,4,5-trisphosphate and inositol
  1,3,4,5-tetrakisphosphate by the oocytes of \textit{Xenopus laevis}.
\newblock {\em J. Biol. Chem.}, 273(7):4052--4058.

\bibitem[Skeel, 1986]{Skeel_MC1986}
Skeel, R.~D. (1986).
\newblock Construction of variable-stepsize multistep formulas.
\newblock {\em Mathematics of Computation}, 47(176):503--510.

\bibitem[Sofroniew and Vinters, 2010]{Sofroniew_AN2010}
Sofroniew, M.~V. and Vinters, H.~V. (2010).
\newblock Astrocytes: biology and pathology.
\newblock {\em Acta Neuropathol.}, 119(1):7--35.

\bibitem[Stryer, 1999]{StryerBiochemistryBOOK}
Stryer, L. (1999).
\newblock {\em Biochemistry}.
\newblock W. H. Freeman and Company, New York, 4th edition.

\bibitem[Suh et~al., 2008]{Suh_BMB2008}
Suh, P.-G., Park, J.-I., Manzoli, L., Cocco, L., Peak, J.~C., Katan, M.,
  Fukami, K., Kataoka, T., Yun, S., and Ryu, S.~H. (2008).
\newblock Multiple roles of phosphoinositide-specific phospholipase~{C}
  isozymes.
\newblock {\em BMB Reports}, 41(6):415--34.

\bibitem[Sun et~al., 2013]{Sun_Science2013}
Sun, W., {McConnell}, E., Pare, J.-F., Xu, Q., Chen, M., Peng, W., Lovatt, D.,
  Han, X., Smith, Y., and Nedergaard, M. (2013).
\newblock Glutamate-dependent neuroglial calcium signaling differs between
  young and adult brain.
\newblock {\em Science}, 339(6116):197--200.

\bibitem[Suzuki et~al., 2004]{Suzuki2004}
Suzuki, Y., Moriyoshi, E., Tsuchiya, D., and Jingami, H. (2004).
\newblock Negative cooperativity of glutamate binding in the dimeric
  metabotropic glutamate receptor subtype {I}.
\newblock {\em J. Biol. Chem.}, 279(34):35526--35534.

\bibitem[Takata et~al., 2011]{Takata_JN2011}
Takata, N., Mishima, T., Hisatsune, C., Nagai, T., Ebisui, E., Mikoshiba, K.,
  and Hirase, H. (2011).
\newblock Astrocyte calcium signaling transforms cholinergic modulation to
  cortical plasticity \textit{in vivo}.
\newblock {\em J. Neurosci.}, 31(49):18155--18165.

\bibitem[Thiel et~al., 1988]{ThielGreengard_PNAS1988}
Thiel, G., Czernik, A.~J., Gorelick, F., Nairn, A.~C., and Greengard, P.
  (1988).
\newblock {Ca$^{2+}$/calmodulin-dependent protein kinase~II: Identification} of
  threonine-286 as the autophosphorylation site in the $\alpha$ subunit
  associated with the generation of {Ca$^{2+}$}-independent activity.
\newblock {\em Proc. Natl. Acad. Sci. USA}, 85:6337--6341.

\bibitem[Togashi et~al., 1997]{TogashiOnaya_BiochemJ1997}
Togashi, S., Takazawa, K., Endo, T., Erneux, C., and Onaya, T. (1997).
\newblock Structural identification of the \textit{myo}-inositol
  1,4,5-trisphosphate-binding domain in rat brain inositol 1,4,5-trisphopshate
  3-kinase.
\newblock {\em Biochem. J.}, 326:221--225.

\bibitem[Vaarmann et~al., 2010]{Vaarmann_JBC2010}
Vaarmann, A., Gandhi, S., and Abramov, A.~Y. (2010).
\newblock Dopamine induces~{Ca$^{2+}$} signaling in astrocytes through reactive
  oxygen species generated by monoamine oxidase.
\newblock {\em Journal of Biological Chemistry}, 285(32):25018--25023.

\bibitem[{van~der~Bend} et~al., 1994]{van-der-Bend1994}
{van~der~Bend}, R.~L., {de~Widt}, J., Hilkmann, H., and Van~Blitterswijk, W.~J.
  (1994).
\newblock Diacylglycerol kinase in receptor-stimulated cells converts its
  substrate in a topologically restricted manner.
\newblock {\em Journal of Biological Chemistry}, 269(6):4098--4102.

\bibitem[Verjans et~al., 1992]{VerjansErneux1992}
Verjans, B., Lecocq, R., Moreau, C., and Erneux, C. (1992).
\newblock Purification of bovine brain inositol-1,4,5-trisphosphate
  5-phosphatase.
\newblock {\em Eur. J. Biochem.}, 204:1083--1087.

\bibitem[Violin et~al., 2014]{Violin_TiPS2014}
Violin, J.~D., Crombie, A.~L., Soergel, D.~G., and Lark, M.~W. (2014).
\newblock Biased ligands at~{G-protein-coupled} receptors: promise and
  progress.
\newblock {\em Trends in Pharmacological Sciences}, 35(7):308--316.

\bibitem[Volterra et~al., 2014]{Volterra_NRN2014}
Volterra, A., Liaudet, N., and Savtchouk, I. (2014).
\newblock Astrocyte {Ca$^{2+}$} signalling: an unexpected complexity.
\newblock {\em Nature Reviews Neuroscience}, 15:327--334.

\bibitem[Walter et~al., 2004]{Walter_JN2004}
Walter, L., Dinh, T., and Stella, N. (2004).
\newblock {ATP} induces a rapid and pronounced increase in
  2-arachidonoylglycerol production by astrocytes, a response limited by
  monoacylglycerol lipase.
\newblock {\em Journal of Neuroscience}, 24(37):8068--8074.

\bibitem[Wang et~al., 2006]{WangNedergaard2006}
Wang, X., Lou, N., Xu, Q., Tian, G.-F., Peng, W.~G., Han, X., Kang, J., Takano,
  T., and Nedergaard, M. (2006).
\newblock Astrocytic {Ca$^{2+}$} signaling evoked by sensory stimulation
  \textit{in vivo}.
\newblock {\em Nat. Neurosci.}, 9(6):816--823.

\bibitem[Weiss, 1997]{Weiss1997}
Weiss, J.~N. (1997).
\newblock The {Hill} equation revisited: uses and misuses.
\newblock {\em FASEB J.}, 11:835--841.

\bibitem[Yamada et~al., 1997]{Yamada_BJ1997}
Yamada, K., Sakane, F., Matsushima, N., and Kanoh, H. (1997).
\newblock {EF}-hand motifs of $\alpha$, $\beta$ and $\gamma$ isoforms of
  diacylglycerol kinase bind calcium with different affinities and
  conformational changes.
\newblock {\em Biochemical Journal}, 321(1):59--64.

\bibitem[Zheng et~al., 2015]{Zheng_Neuron2015}
Zheng, K., Bard, L., Reynolds, J.~P., King, C., Jensen, T.~P., Gourine, A.~V.,
  and Rusakov, D.~A. (2015).
\newblock Time-resolved imaging reveals heterogeneous landscapes of nanomolar
  {Ca$^{2+}$} in neurons and astroglia.
\newblock {\em Neuron}, 88(2):277--288.

\bibitem[Zorec et~al., 2012]{Zorec_ASN2012}
Zorec, R., Araque, A., Carmignoto, G., Haydon, P., Verkhratsky, A., and
  Parpura, V. (2012).
\newblock Astroglial excitability and gliotransmission: {An appraisal of
  Ca$^{2+}$ as a signaling route}.
\newblock {\em ASN Neuro}, 4(2):e00080.

\end{thebibliography}

\end{document}